%% file: stasinska.tex
\documentclass[oldversion]{aa}
\usepackage{graphicx}
\usepackage{lscape}
\usepackage{longtable}

\usepackage{txfonts}
\usepackage{natbib}\bibpunct{(}{)}{;}{a}{}{,} 

\newcommand{\msun}{M$_{\odot}$}

\newcommand{\lsun}{L$_{\odot}$}

\newcommand{\sbs}{SBS\,1150+599A}
\newcommand{\png}{PN\,G\,135.9+55.9}
\newcommand{\TS}{TS\,01}

\newcommand{\cmcub}{~cm$^{-3}$}
\newcommand{\kms}{~km~s$^{-1}$}

\newcommand{\ergcmsq}{~erg~cm$^{-2}$~s$^{-1}$}

\newcommand{\Ha}{H$\alpha$}
\newcommand{\Hb}{\ifmmode {\rm H}\beta \else H$\beta$\fi}
\newcommand{\hi}{H~{\sc i}}

\newcommand{\Hei}{He~{\sc i} $\lambda$5876}
\newcommand{\heii}{He~{\sc ii}}
\newcommand{\Heii}{He~{\sc ii} $\lambda$4686}
\newcommand{\Heiiuv}{He~{\sc ii} $\lambda$1640}

\newcommand{\Civ}{C~{\sc iv} $\lambda$1549}

\newcommand{\Oiii}{[O~{\sc iii}] $\lambda$5007}

\newcommand{\Neiii}{[Ne~{\sc iii}] $\lambda$3869}
\newcommand{\neiii}{[Ne~{\sc iii}]}
\newcommand{\nev}{[Ne~{\sc v}]}
\newcommand{\Nev}{[Ne~{\sc v}] $\lambda$3426}

\newcommand{\Oivir}{[O\,{\sc{iv}}]\,$\lambda$25.9\,$\mu$m}
\newcommand{\Nevir}{[Ne\,{\sc{v}}]\,$\lambda$24.3\,$\mu$m}

\usepackage{color}

\begin{document}
   \title{The chemical composition of \TS, the most oxygen-deficient planetary nebula}
\subtitle{AGB nucleosynthesis in a metal-poor binary star\thanks{Based on observations obtained at the Canada-France-Hawaii Telescope (CFHT)
which is operated by the National Research Council of Canada, the Institut
National des Sciences de l'Univers of the Centre National de la Recherche
Scientifique of France, and the University of Hawaii}$^,$ \thanks {Based on observations with the NASA/ESA Hubble Space Telescope, obtained at the Space Telescope Science Institute, which is operated by the Association of Universities for Research in Astronomy, Inc., under NASA contract NAS 5-26555.}$^,$ \thanks{Based on observations made with the Spitzer Space Telescope,
which is operated by the Jet Propulsion Laboratory, California Institute
of Technology, under NASA contract 1407.}}

   \author{G. Stasi\'nska\inst{1}
          \and C. Morisset\inst{2}
             \and G. Tovmassian\inst{3}
               \and T. Rauch\inst{4}
                \and M. G. Richer\inst{3}
                   \and M. Pe\~na\inst{2}
                      \and R. Szczerba\inst{5}
                            \and T. Decressin\inst{6}
                          \and C. Charbonnel\inst{7}
                             \and L. Yungelson\inst{8}
                          \and R. Napiwotzki\inst{9}
                         \and S. Sim\'on-D\'iaz\inst{10}
                            \and L. Jamet\inst{1}                            
          }
   \offprints{G. Stasi\'nska}
   \institute{LUTH, Observatoire de Paris, CNRS, Universit\'e Paris Diderot; Place Jules Janssen 92190 Meudon, France\\
\email{grazyna.stasinska@obspm.fr}
\and
Instituto de Astronomia, Universidad Nacional Autonoma de Mexico, Apdo. Postal 70264, Mexico D.F., 04510 Mexico
\and
Instituto de Astronomia, Universidad Nacional Autonoma
de Mexico, Apdo. Postal 877, Ensenada, Baja California, 22800 Mexico
\and
Institute for Astronomy and Astrophysics,
Kepler Center for Astro and Particle Physics,
Eberhard Karls University, Sand 1, 
72076 T\"ubingen,
Germany
\and
N. Copernicus Astronomical Center, Rabia\'nska 8, 87-100 Toru\'n, Poland
\and
Argelander Institute for Astronomy (AIfA), Auf dem H\"ugel 71, D-53121 Bonn,
Germany
\and
Geneva Observatory, University of Geneva, ch. des Maillettes 51, 1290 Sauverny, Switzerland and Laboratoire d'Astrophysique de Toulouse-Tarbes,
  CNRS UMR 5572, Universit\'e de Toulouse, 14, Av. E.Belin, F-31400 Toulouse, France
\and
Institute of Astronomy of the Russian Academy of Sciences, 48 Pyatniskaya Str., 119017 Moscow, Russia
\and
Centre for Astrophysics Research, University of Hertfordshire, College Lane, HatÞeld 
AL109AB, UK 
         \and
Instituto de Astrof\'isica de Canarias, E38200, La Laguna, Tenerife, Spain
}

   \date{ }

 
  \abstract
{The planetary nebula \TS\ (also called \png\ or \sbs), with its record-holding low oxygen abundance and its double degenerate close binary core (period 3.9 h),  is an exceptional object located in the Galactic halo.

 We have secured observational data in a complete wavelength range in order to pin down the abundances of half a dozen elements in the nebula.  The abundances are obtained via detailed photoionization modelling taking into account all the observational constraints (including geometry and aperture effects) using the pseudo-3D photoionization code Cloudy$\_$3D. The spectral energy distribution of the ionizing radiation is taken from appropriate model atmospheres. Incidentally, from the new observational constraints, we find that both stellar components  contribute to the ionization: the ``cool'' one provides the bulk of hydrogen ionization, and the ``hot'' one  is responsible for the presence of the most highly charged ions, which explains why previous attempts to model the nebula experienced difficulties. 
   
   The nebular abundances of C, N, O, and Ne are found to be respectively,  1/3.5, 1/4.2, 1/70, and   1/11 of the Solar value, with uncertainties of a factor 2. Thus the extreme O deficiency of this object is confirmed. The abundances of S and Ar are less than  1/30 of  Solar.  The  abundance of He relative to H is  0.089$\pm$0.009. 
   
Standard models of stellar evolution and nucleosynthesis  cannot explain the  abundance pattern observed in the nebula.  To obtain an extreme oxygen deficiency in a star whose progenitor has an initial mass of about 1\,\msun\   requires an additional mixing process, which can be induced by stellar rotation and/or  by the presence of the close companion.  We have computed a  stellar model with initial mass of 1\,\msun, appropriate metallicity, and initial rotation of 100\kms, and find that rotation greatly improves the agreement between  the predicted and observed abundances.
}


   \maketitle
%

\section{Introduction}
\label{sec:introduction}

\sbs\ was discovered in the second Byurakan Sky Survey  and first classified as a cataclysmic variable \citep{1999PASP..111.1099S}. \cite{2001A&A...370..456T}  discussed in detail the nature of the object and came to the conclusion that, in fact, it is a  planetary nebula (PN). The  object was renamed  \png, following the nomenclature for Galactic PNe from the Strasbourg-ESO catalogue of Galactic Planetary Nebulae \citep{1992secg.book.....A}.  For the sake of brevity, we will refer to it as \TS\ in the rest of the paper. This PN is special in at least three important aspects.  First of all, its oxygen abundance is very low, significantly lower than in any other PN known up to now \citep{2001A&A...370..456T, 2002A&A...395..929R, 2002AJ....124.3340J, 2005A&A...430..187P}. Second, its nucleus is a spectroscopic binary, with a  period  of only a few hours \citep{2004ApJ...616..485T}. Third, it appears, from estimates of the nature and masses of the two stellar components, that \TS\  could turn into a double degenerate Type Ia Supernova \citep{2004ApJ...616..485T}. Each of these aspects, even taken alone, makes  \TS\ an exceptional object.

In this paper, we reexamine the  chemical composition of \TS.
Briefly, the story of the determination of the chemical composition of this object is the following. Tovmassian et al. (2001) had optical spectra of 
\TS\ in the range 3900--7000\,\AA\  obtained with 2\,m class telescopes which showed no lines from heavy elements except a very weak \Oiii, with an intensity a few percent of H$\beta$. A coarse photoionization  analysis  suggested an oxygen abundance smaller than 1/100 of Solar. Note that standard empirical methods for abundance determinations in PNe cannot be used for \TS, since the electron temperature cannot be determined directly from observations. To go further  in the abundance determination of \TS\ required an estimate of the  effective temperature of the central star. One way is to obtain a good blue spectrum of the PN, and use  the   \Nev/ \Neiii\ ratio (or a limit on it) as a constraint.  \cite{2002A&A...395..929R} at the Canada-France-Hawaii Telescope (CFHT) and \cite{2002AJ....124.3340J} at the Multiple Mirror Telescope (MMT) secured deep blue spectra in order to detect these lines. \cite{2002AJ....124.3340J} detected the \Nev\ line at a level of ~0.8 H$\beta$.  \cite{2002A&A...395..929R} found only an upper limit of 0.1 H$\beta$! Concerning  the \Neiii\ line, {\cite{2002AJ....124.3340J} measured an intensity about 10 times larger than \cite{2002A&A...395..929R}. The two papers appeared within a few days of each other on astro-ph, revealing this big conflict in the observations. The two groups conducted independent photoionization analyses, and both concluded that the O/H ratio is less than 1/100 of Solar (the main reason for their similar result for the oxygen abundance was the similar \Nev/\Neiii\ ratio used by both studies).  \cite{2005A&A...430..187P} merged and discussed the two observational data sets and conducted their own photoionization analysis. They concluded that the O/H ratio of \TS\ lies between 1/30 - 1/15 of Solar (still holding the record for the most oxygen poor planetary nebula but much higher than previously published). However,  \cite{2005A&A...430..187P} neglected to consider observations of \TS\ made with the \textit{Hubble Space Telescope} (HST) and  the \textit{Far Ultraviolet Spectroscopic Explorer} (FUSE). As a result, some of their ``predicted'' line intensities  are in conflict with what is actually observed in the UV. HST observations were obtained in 2003, and presented in a short, preliminary version by {\cite{2006IAUS..234..431J}. Those authors quoted an oxygen abundance of 1/30 - 1/40 of Solar, and carbon and nitrogen abundances roughly 1/10 of Solar.

Before embarking on a new determination of abundances, we have chosen to gather the best possible observations at all wavelength ranges. These data provide many more constraints than were available in any previous study. In order to make the best use of the large amount of data obtained with different telescopes, we use a pseudo-3D photoionization code, Cloudy\_3D, which is able to account for the nebular geometry as we see it now, and with which we can properly take into account the aperture effects. This code is based on CLOUDY \citep{1998PASP..110..761F} and was written by \cite{2006IAUS..234..467M}. 

The paper is organized as follows. Section 2  presents the new observational material: several optical spectra, HST imaging and spectroscopy, infrared spectroscopy with the \emph{Spitzer} Telescope, and mentions our X-ray observations with XMM. Section 3 summarizes other data that we  used as constraints for the photoionization modelling. Section 4 describes our modelling strategy, and presents our ``reference model''. Section 5 evaluates the error bars on the derived elemental abundances, taking into account  observational uncertainties in emission-line fluxes, uncertainties in model input parameters as well as uncertainties arising from an imperfect description of the physical processes included in the models. In Section 6, we compare the chemical composition of \TS\ with that of other PNe in the Galactic halo and discuss it in terms of stellar  nucleosynthesis in the Asymptotic Giant Branch (AGB) phase. Finally, Section 7 summarizes our main findings.

\section{New observational data on emission lines}
\label{sec:observations}

We present the observational data that  we secured on \TS\ and its stellar core since the work presented in Tovmassian et al. (2004). Some of those data were already briefly reported in conference proceedings, but here we describe the acquisition and reduction processes in more detail. Note that the observations and reductions were done by different people and at different epochs, when our knowledge on the object was not the same. This explains the differences in the tactics employed to reduce the data, estimate the line fluxes and correct for reddening. We did not try to fully homogenize the data reduction process, since we felt it unnecessary for our purposes.

The log and characteristics of each set of observations are given in Table \ref{tab:log}.

\begin{table} [hb!]
\caption{Log and characteristics of the spectroscopic observations}  
\label{tab:log} 
\scalebox{0.95}{                    
\begin{tabular}{llllc}       
\hline\hline 
telescope     & date & $\lambda$ range & resolution & aperture  \\
\hline
FUSE & 30 Jan 02 & 900--1200\,\AA &   & $30 \times 30$\,\arcsec \\
HST STIS  & 4 May 03 & 1170--1700\,\AA &
1.20\,\AA & 0.5\,\arcsec \\
CFHT MOS & 1 May 03 & 3400--5300\,\AA  & 3.0--3.5\AA &  1\,\arcsec \\
CFHT MOS & 4 Mar 01 & 3400--8000\,\AA\ &
23\AA &  5\,\arcsec \\
Kitt Peak & 1 Jan 03 & 3600--7500\,\AA &  7\AA  & 2\,\arcsec \\
SDSS &  &    3819--9196\,\AA  & 2--4\AA  & diameter 3\,\arcsec  \\
Spitzer IRS SH  & 22 Apr 06 & 9.9--19.6\,$\mu$m & 600 & $4.7 \times 11.3$\,\arcsec \\
Spitzer IRS LH   & 22 Apr 06 & 18.7--37.2\,$\mu$m  & 600 & $11.1 \times 22.3$\,\arcsec\\
\hline                                   
\end{tabular}
}
\end{table}

\subsection{Imaging}
\label{sec:imaging}

We (M.P.) retrieved from the HST archives and analyzed the data corresponding to the  proposal ID 9466.   The observations were performed on May 5, 2003. Two types of data are available: 
direct imaging and spectroscopy.

Direct imaging was obtained with the Advanced Camera for Surveys (ACS). 
The High Resolution Channel with a field of view of 26\arcsec$\times$29\arcsec
and a plate-scale of 0.027\arcsec per pixel, with filters around H$\alpha$
(central wavelength 6581.97 $\pm$ 162.8\,\AA) and \Nev\ (central wavelength
3432.8 $\pm$ 42.66\,\AA) were used.

\begin{figure} [h!]
\includegraphics[width=9cm]{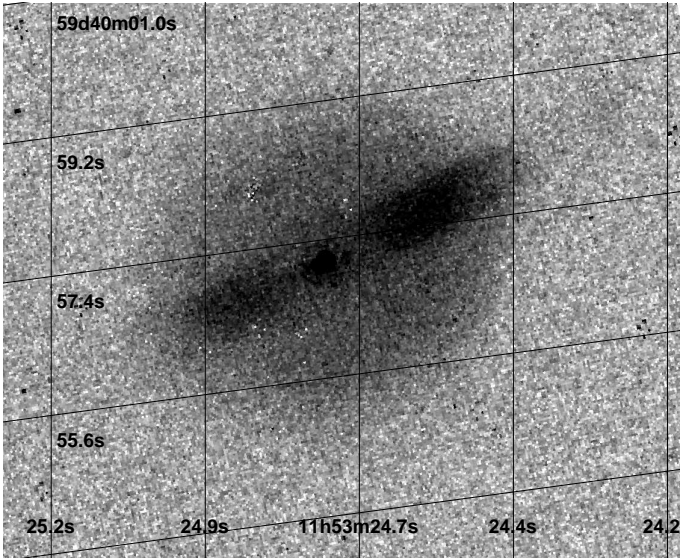}
  \caption{HST-ACS  H$\alpha$ image of \TS, with a logarithmic grey scale.  A grid with the position scale (R.A.  and Dec) is traced. North is up.}
    \label{fig:Ha-image}
\end{figure}

Figure  \ref{fig:Ha-image} shows the H$\alpha$ image obtained by averaging the 4 calibrated frames:  j8do01021, j8do01022, j8do01023, j8do01024 (870 s exposure time each; 58 min in total), after aligning them with respect to j8do01021.   The image is roughly elliptical in shape with two brighter,
symmetrically-placed lobes at a position angle of about $103^{\circ}$
that extend the full major axis of the ellipse. The nebula is not perfectly symmetric, with the outermost southern  part much fainter. The size of the nebular image is about 5\arcsec.

\begin{figure}
\includegraphics[width=9cm]{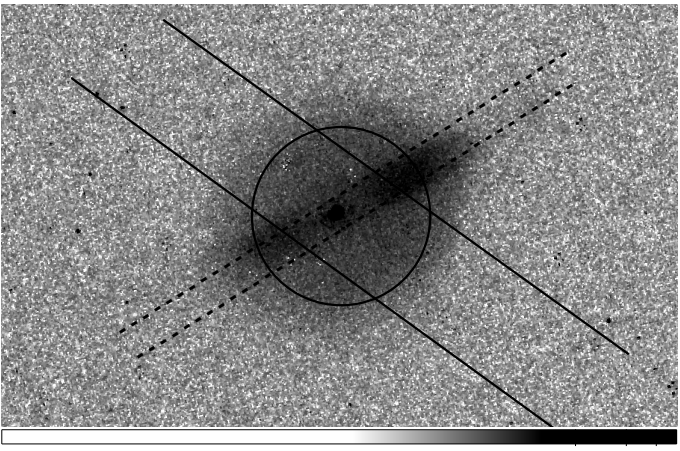}
  \caption{HST-ACS  H$\alpha$ image of \TS. The locations of the different spectroscopic apertures are indicated: Kitt Peak (continuous lines), HST STIS (dotted lines), SDSS (circle). See text for the CFHT 2003 observations.}
    \label{fig:Ha_slits}
\end{figure}

Figure \ref{fig:Ha_slits} shows the same image as Fig. \ref{fig:Ha-image}, with the different observing apertures indicated: continuous lines for Kitt Peak, dotted lines for HST STIS, circle for SDSS. For the CFHT 2003 observations, the slit was rotated before each of the seven exposures (see Sect. \ref{sec:cfht-data}), so as to remain as close as possible to the parallactic angle. The position of the slit is not indicated in the figure, for the sake of clarity, but was taken into account correctly when comparing the predicted line intensities  with the observed ones (see Sect. \ref{sec:fluxes-scale}).

\begin{figure}
\includegraphics[width=9cm]{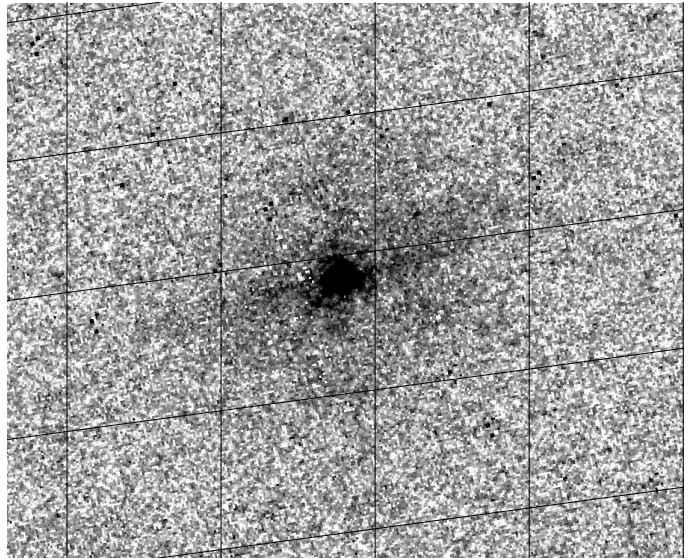}
  \caption{HST-ACS   image of \TS\ in the \Nev\ line, on a logarithmic grey scale. Same orientation and scale as Fig. \ref{fig:Ha-image}. }
    \label{fig:SBS_3433}
\end{figure}

Figure \ref{fig:SBS_3433} shows an average of the \nev $\lambda$3432 calibrated images
j8do01011, j8do01012, j8do01013 and j8do01014 (3000 sec of exposure time each;
200 minutes in total).    This image reveals a very faint,  roughly spherical extended
 nebulosity and an important emission in the center, probably  caused by 
the stellar core.   Some faint extensions are marginally detected in the directions of the  
H$\alpha$ lobes.

\subsection{Optical spectroscopy}
\label{sec:optical-spectroscopy}

\subsubsection{CFHT  data }
\label{sec:cfht-data}

\paragraph{\textbf{CFHT  2003}}
\label{sec:cfht-2003}

\TS\ was reobserved at the CFHT by M.R. and G.S. on  1 May 2003 using the MOS spectrograph and a 1\arcsec\ slit \citep{1994A&A...282..325L}.  The U900 grism was used, giving a spectral range of  3400--5300\,\AA\  and a spectral resolution of 3--3.5\,\AA\ (measured from arc lamp spectra).  Seven 1800\,s exposures were obtained.  During each exposure, the slit was set to within 10$^\circ$ of the parallactic angle. Details of the reduction process of the individual exposures are given in Tovmassian et al. (2004).  

To obtain a high signal-to-noise spectrum of the nebular emission lines, it is necessary to account for the stellar and nebular continuum emission.  These contributions were subtracted from the individual exposures before summing the individual spectra.  First, the observed spectra were shifted in velocity so that the stellar absorption line was at rest.  Next, the H$\beta$ intensity was measured and used to scale a model of the nebular continuum emission.  Since the H$\beta$ flux is affected by stellar absorption, the stellar absorption was assumed to have an equivalent width of 13\,\AA, a value typical for the models used (see below).  This scaled nebular continuum was then subtracted from the observed spectrum.  Then, a model stellar atmosphere  was scaled so as to match the observed continuum level and subtracted from the observed spectrum.  This procedure leaves a pure nebular emission line spectrum, supposing that the model nebular and stellar continua are representative of their real counterparts.  It is unlikely that subtracting the continua introduces significant uncertainty into our final line intensities.  Model stellar spectra with $(T_{eff}, \log g)$ pairs of (90\,kK, 5.05), (120\,kK, 5.35), and (150\,kK, 5.56) computed by R.N. (see Tovmassian et al. 2004) were subtracted from our observed spectra and the differences in the resulting line intensities were always smaller than the uncertainties in the fits\footnote{This treatment was applied well before we had understood that the optical continuum was dominated by a star of 55\,kK (see Sect. \ref{sec:stellar-core}). In view of the fact that the lines we use for the dignostics discussed in the present paper are hardly affected by this correction, we decided not to redo the subtraction using more adequate model stellar spectra.}.  

Once the stellar and nebular continua were subtracted, the individual nebular spectra were shifted back to their original velocities and summed.  We measured the nebular emission line strengths from this final spectrum.  The line intensities were measured using INTENS, a locally-implemented software package \citep{1985ApJS...57....1M}.  This software simultaneously fits a sampled Gaussian function to the emission line(s) and a straight line to the continuum.  It returns the line strengths, line wavelengths, and uncertainties in these quantities.  The line intensities presented in Table 2, together with their uncertainties,  are those measured after subtracting the stellar spectrum for $T_{\rm eff} = 120$k\,K and $\log g = 5.35$. The listed intensities are not corrected for reddening. In the case of no detection, two-sigma upper limits are given instead.

We note that the \Nev\ line is present, and strong. Its intensity is of the same order as found in the spectrum of \cite{2002AJ....124.3340J}, and much higher than the upper limit given by \cite{2002A&A...395..929R}. The remaining lines have intensities roughly in agreement with those published by \cite{2002A&A...395..929R} and \cite{2002AJ....124.3340J}, except for the \neiii\ line which appeared on the top of a bump in \cite{2002AJ....124.3340J} and was attributed a high intensity in that paper. 
\begin{figure}
\includegraphics[width=9cm]{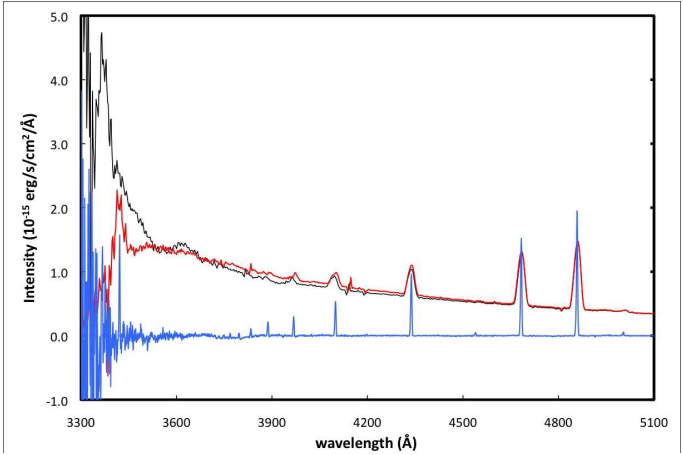}
  \caption{Flux-calibrated CFHT spectra of \TS.  In blue is the CFHT 2003
spectrum \emph{with the nebular and stellar continua subtracted}, in black  the
original CFHT 2001 spectrum, and in red the CFHT 2001 spectrum with the
correct wavelength solution.}
    \label{fig:cfh_spec}
\end{figure}

\paragraph{\textbf{CFHT  2001}}
\label{sec:cfht-2001}

In view of the large discrepancy with the \cite{2002A&A...395..929R} data concerning the \Nev\ line, we decided to reanalyze the spectrum of \TS\ we had obtained in March 2001 at the CFHT.  First, though, we refer the interested reader to Richer et al. (2002) for a discussion of the details of these observations.  The basic difficulty with these observations was that the arc lamp spectra were taken with the same 5\arcsec slit used for the object observations, which resulted in an arc spectrum with severely-blended lines.  After repeating the wavelength calibration more carefully, we found that our previous solution had stretched out the spectrum at the shortest wavelengths, leading us to not recognize the \Nev\ line because of its low contrast with respect to the continuum and its erroneous wavelength.  The analysis of this spectrum was considerably simpler given that the stellar features were not resolved.  We simply fit the continuum shape and removed it (INTENS assumes that the continuum is a straight line), then measured the line intensities with INTENS.  The resulting line intensities are given in Table 2. The intensity we find now for \Nev\ is in agreement with the one given by  \cite{2002AJ....124.3340J}, and compatible within 2 sigmas with the one obtained with the CFHT 2003 data. Slit effects could perhaps explain the slight difference in intensity between CFHT 2001 and CFHT 2003.

\subsubsection{Kitt Peak data }
\label{sec:kittpeak-data}

 \TS\ Êwas observed at the 4\,m telescope of Kitt Peak National Observatory on 1 January 2003. The grating used was KPC-10A, and the slit 2\arcsec$\times$300\arcsec, with an orientation of P.A.=44.7$^\circ$. Two exposures of 600\,s were obtained. The data were reduced by L.J., employing the same procedure as for the SDSS spectrum, explained below.

\subsubsection{SDSS and Kitt Peak data }
\label{sec:sdss-data}

The spectrum of \TS\ appears in the data of the Sloan Digital Sky Survey SDSS  (http://www.sdss.org) under the name 0953-52411-160\footnote{The analysis by L.J. was done soon after our discovery of the spectrum in SDSS data release 2. In data release 6 \citep{2008ApJS..175..297A}, SDSS spectra were recalibrated, resulting in an increase of about 30\% of the fluxes of \TS. Line ratios remained unchanged. Therefore, the analysis of L.J. remains valid. On the other hand, whenever we needed to consider the \TS\ continuum in this paper, we used the recalibrated spectrum.}. We present here its analysis as performed by  L.J.

We separated the nebular emission from the stellar spectrum and evaluated the reddening with as few free parameters as possible. We assumed the stellar spectrum to be that of a single white dwarf (WD), hence neglecting the possible contribution of the companion; we considered three model WD spectra at temperatures of 90, 120 and 150\,kK (the same as used for the CFHT 2003 spectrum). As for the nebular continuum, we computed the free-free and free-bound emissivities of H$^+$ and He$^{++}$ with the {\sc chianti} code \citep{2006ApJS..162..261L}, assuming an abundance ratio He$^{++}$/H$^+=0.075$ and an electronic temperature of 30\,kK (Richer et al. 2002). We also retrieved the H$\gamma$ nebular emissivity at this temperature from \cite{1995MNRAS.272...41S}.

First, we computed a model of the total (stellar+nebular) spectrum around the H$\gamma$ line. In each of the WD spectral models, the H$\gamma$ line has a Voigt profile with a given equivalent width (EW) $W_{\rm wd}$, Gaussian width $\sigma_{\rm wd}$ and Lorentzian width $a_{\rm wd}$. As for the nebular emission, we computed the EW $W_{\rm neb}$ of the emission line with respect to the nebular continuum. Furthermore, we assumed the real width of the nebular line to be much smaller than the instrumental one, so its observed Gaussian and Lorentzian widths, respectively $\sigma_{\rm inst}$ and $a_{\rm inst}$, are representative of the instrumental PSF. Consequently, the observed widths of the stellar line are $\sigma_{*}=(\sigma_{\rm wd}^{2}+\sigma_{\rm inst}^{2})^{1/2}$ and $a_{*}=a_{\rm wd}+a_{\rm inst}$. We normalized the local stellar+nebular continuum with the fit of a slope on either side of the line. Finally, we let the central wavelengths of the stellar and nebular line, respectively $\lambda_{*}$ and $\lambda_{\rm neb}$, be independent from each other. Using the data and assumptions gathered, we fitted a consistent model on the observed spectrum around the line. Calling $V_{\rm neb}(\lambda)$ the profile of the nebular line, $V_{*}(\lambda)$ that of the star (both being normalized to an EW of unity), and $C_{*}$ the stellar contribution to the flux, the model can be written as
\begin{eqnarray}
F_{\lambda}(\lambda)&=&C_{*}\,\left(1-W_{\rm wd} V_{*}(\lambda-\lambda_{*};\,\sigma_{\rm inst}, a_{\rm inst}, \sigma_{\rm wd}, a_{\rm wd})\right)+\nonumber\\
&&+(1-C_{*})\,\left(1+W_{\rm neb} V_{\rm neb}(\lambda-\lambda_{\rm neb};\,\sigma_{\rm inst}, a_{\rm inst})\right)
\end{eqnarray}
For all three WD model atmospheres, we obtained very good fits with no visible systematic residuals.

Given that the intrinsic H$\alpha$/H$\beta$ nebular line ratio in \TS\ is not merely the case B recombination value, and given the additional problems with H$\alpha$ (see Sect. \ref{sec:Ha}), we cannot use this ratio to evaluate the extinction. Hence, we used the continuum to measure the latter. We first corrected the data for the small foreground extinction ($E(B-V)=0.029$) estimated by \cite{1998ApJ...500..525S}. Then, we removed most of the nebular or stellar lines from the observations making use of a median filter. Finally, adopting the SMC extinction law \citep{1984A&A...132..389P, 1985A&A...149..330B} and comparing the filtered spectrum with the theoretical continuum, we evaluated the reddening and corrected the data for it. The choice of the extinction law was motivated by the low metallicity of \TS. The reddening amounts obtained (additionally to the foreground one) are $E(B-V)=0.033$, 0.044 and 0.050 for the 90, 120 and 150\,kK WD models, respectively.

The last processing of the data was the removal of the stellar and nebular continua, in particular to avoid the contamination of the nebular lines by the underlying stellar features. We used the fit of the H$\gamma$ line to shift the theoretical stellar spectrum and nebular continuum according to their evaluated radial velocities, convolved them by the average instrumental PSF and subtracted them. Finally, we identified visually the detectable lines and measured their fluxes. The fully processed SDSS spectrum where the 90\,kK WD model spectrum was removed is presented in Fig.~\ref{fig:sloan_all}, while Fig.~\ref{fig:izotov_all} shows the  result of the processing of the Kitt Peak spectrum. The choice of the WD model had a moderate impact on the evaluation of the line fluxes, of order of $2$\% for most of them and $2\sigma$ in the worst case. The intensities of the SDSS spectrum and the Kitt Peak spectrum are listed in Table \ref{tab:optical-lines}.

\begin{figure}
\includegraphics[width=9cm]{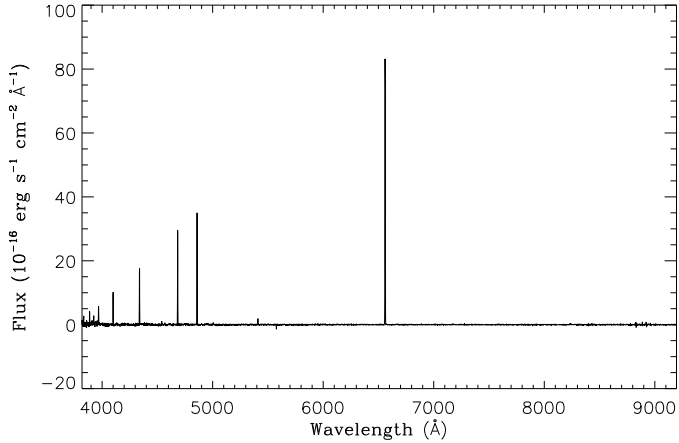}
  \caption{Fully processed SDSS spectrum (dereddened, free of continuum and stellar lines).}
    \label{fig:sloan_all}
\end{figure}

\begin{figure}
\includegraphics[width= 9cm]{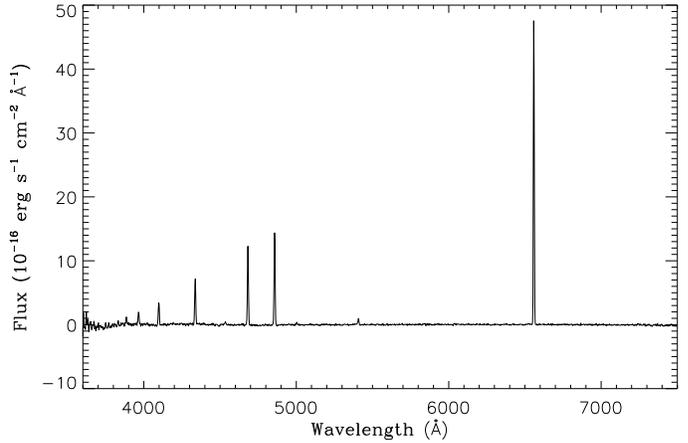}
  \caption{Fully processed Kitt Peak spectrum (dereddened, free of continuum and stellar lines).}
    \label{fig:izotov_all}
\end{figure}

\begin{table} [ht]
\caption{Intensities of optical lines, corrected for stellar absorption, but not for reddening, with respect to H$\beta$=100}
\begin{center}
\scalebox{0.80}{
\begin{tabular}{rrrrrr}       
\hline\hline 
\input{SBSallfluxes-paper.txt}
\hline 
\end{tabular}}
\label{tab:optical-lines} 
\end{center}
\end{table}
\normalsize

\subsection{Ultraviolet spectroscopy}
\label{sec:ultr-spectr}

\subsubsection{HST STIS data}
\label{sec:hst-stis-data}

The HST STIS spectroscopic data correspond to the same  proposal (ID 9466) as the imaging data. 

A 52 \arcsec\ $\times$ 0.5 \arcsec\ slit was used.  It was oriented along
the bright jet-like emission of the nebula (P.A. 103$^\circ$),  see Fig. \ref{fig:Ha_slits}.

\paragraph{\textbf{Far UV observations}}
\label{sec:far-uv-observations}

\begin{figure} [h]
\includegraphics[width=9cm]{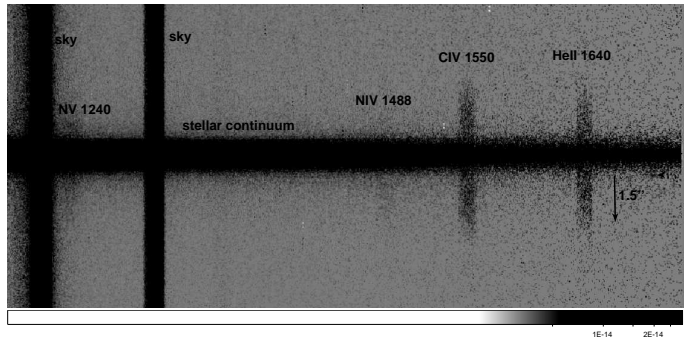}
  \caption{2D calibrated HST FUV spectrum of \TS}
    \label{fig:FUV-2D}
\end{figure}

  The MAMA detector combined with a G140L grating provided 2D spectra o8do03020,
o8do03030, o8do03040, o8do03050 and o8do03060, with 4675 secs exposure time each. The spectra cover a wavelength range from 1170 to 1700\,\AA, with a resolving power of 1190
at the central wavelength 1425\,\AA. 
The 2D spectra show a bright blue stellar continuum and a few faint and extended
emission lines from the nebula.  Calibrated 2D spectra were combined (after
shifting because the spectroscopic observations were dithered) to produce a 389.6 min 
spectrum.  The resulting 2D spectrum is shown in Fig. \ref{fig:FUV-2D}.  
The stellar spectrum shows good signal-to-noise and stellar and
interstellar absorption are present.   The analysis of the stellar spectrum is presented in a companion paper \citep{Tovmassian09}. Regarding  nebular lines, the following ones  are detected:    N\,{\sc{v}}\,$\lambda$1240,  N\,{\sc{iv}}]\,$\lambda$1488,  C\,{\sc{iv}}]\,$\lambda$1550 and  He\,{\sc{ii}}]\,$\lambda$1640. Selective absorption of resonance lines by the intervening interstellar medium is treated in Sect. \ref{sec:interstellar_abs}.

\begin{figure} [h]
\includegraphics[width=9cm]{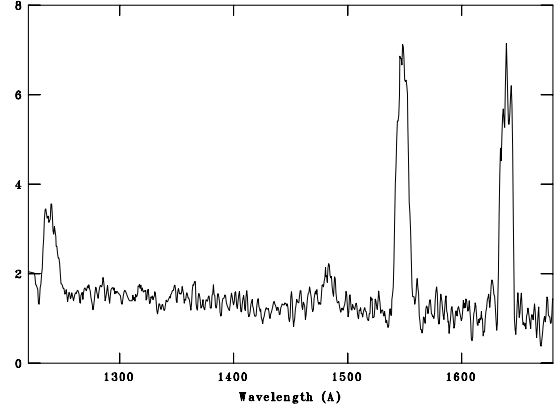}
  \caption{HST FUV nebular spectrum spectrum of \TS, not corrected for reddening, showing the  He\,{\sc{ii}}\,$\lambda$1640\,\AA, N\,{\sc{v}}\,$\lambda$1240\,\AA, N\,{\sc{iv}}]\,$\lambda$1488\,\AA, and  
 C\,{\sc{iv}}\,$\lambda$1550\,\AA\ lines. The fluxes are in $10^{-14}$ erg cm$^{-2}$ s$^{-1}$\AA$^{-1}$.}
      \label{fig:FUVneb}    
\end{figure} 
From the combined 2D spectrum the nebular emission was extracted on both sides
of the central star, with an extraction window of 60 pixels, equivalent to
1.464\arcsec.  Fig. \ref{fig:FUVneb}  is a combination of both nebular spectra. The line fluxes in each lobe, and the combined values with respect  He\,{\sc{ii}}]\,$\lambda$1640 are listed in Table \ref{tab:HST-int2}
.

\begin{table}
\caption{Observed HST UV line fluxes, relative to He II 1640}  
\label{tab:HST-int2} 
\scalebox{0.95}{
\begin{tabular}{lccc}       
\hline\hline 
 ion lambda  &  bright lobe  &  faint lobe  &  combined  \\
\hline 
   He\,{\sc{ii}}\,$\lambda$1640\,\AA   &   1.56$^a$  &   1.15$^a$    &  (2.7 +/- 0.3)$^a$ $^b$ \\
 \hline 
   N\,{\sc{v}}\,$\lambda$1240\,\AA     &   0.31  &       0.51      &       0.47 +/- 0.04 \\
   O\,{\sc{iv}}\,$\lambda$1402\,\AA       &  $<$ 0.06 &        ---     &     $< 0.06$   \\
   N\,{\sc{iv}}]\,$\lambda$1488\,\AA      & 0.17       &  noisy       &    0.12 +/- 0.05 \\
   C\,{\sc{iv}}\,$\lambda$1550\,\AA      &  1.10       &  1.33      &       1.28 +/- 0.10 \\
\hline 
   N\,{\sc{iii}}]\,$\lambda$1750\,\AA &   $<$ 0.1       &           &           $<$ 0.1 \\
   C\,{\sc{iii}}]\,$\lambda$1909\,\AA   &  $<$ 0.1       &           &            $<$ 0.1 \\

\hline                             
\multicolumn{4}{l}
{$^a$ The flux of  He\,{\sc{ii}}\,$\lambda$1640 is in units of 10$^{-14}$ \ergcmsq} \\
\multicolumn{4}{l}
{$^b$ All the flux in  He\,{\sc{ii}}\,$\lambda$1640, including both lobes} \\
\hline
\end{tabular}}
\end{table}

\paragraph{\textbf{Near UV observations}}
\label{sec:near-uv-observations}


The MAMA detector combined with a G230LL grating provided 2D spectra o8do02010,
o8do02020, o8do02030, o8do02040 and o8do02050, covering a wavelength range
from about 1600 to 3150\,\AA.  The calibrated 2D spectra were combined (after
aligning) to produce a spectrum with total exposure time of  237.5  min.    As for the FUV, the NUV  stellar spectrum has good signal-to-noise and stellar and
interstellar absorption can be seen.  However, no  nebular lines are detected. In particular, \Heiiuv,  N\,{\sc{iii}}]\,$\lambda$1750, and  C\,{\sc{iii}}]\,$\lambda$1909 are not seen.  Table \ref{tab:HST-int2} gives upper limits for the latter line intensities, with respect to  \Heiiuv, as seen in the FUV spectrum.

\subsubsection{FUSE data}
\label{sec:fuse-data}
The observations of \TS\ with the \emph{Far Ultraviolet Spectroscopic Explorer (FUSE)} and their reductions (done by G.T.) are described in \cite{2004ApJ...616..485T}. No emission lines were detected in the observed wavelength region between 900 and 1200\,\AA, except the H Ly$\beta$ line. 
For the photoionization modelling of the nebula, it is important to determine upper limits to the intensities of nebular lines expected in this wavelength range. 
We proceeded in the following way. From a previous model of \TS\  we took the computed nebular continuum.  We superimposed on it the lines  C\,{\sc{iii}}\,$\lambda$977.020,  N\,{\sc{iii}}\,$\lambda$989.799 and  He\,{\sc{ii}}\,$\lambda$992.4
with FWHM of 0.1\,\AA\ (which corresponds to the measured expansion velocity 30 km sec$^{-1}$, see Sect. \ref{sec:expansion-velocity}). To this we added the central star model mentioned in Sect. \ref{sec:cfht-2003}.
The resulting spectrum was processed through the interstellar hydrogen
absorption simulator (http://violet.pha.jhu.edu/~gak/fwebsim.html) to be
compared with the observations.
It turned out that the corresponding lines start to be detectable in the
resulting spectrum when the line flux reaches approximately $7.5 \times 10^{-14}$ ergs cm$^{-2}$ s$^{-1}$.
Indeed, the wavelength region that we are exploiting here is very complicated.
Apart from the different interstellar absorptions and terrestrial airglow,
the lines in this region also lie at the edges of the detectors where they overlap and
errors are much higher compared to  other regions to the red.

\subsection{Mid-infrared spectroscopy}
\label{sec:mid-infr-spectr}

\begin{figure} [h]
\includegraphics[width=9cm]{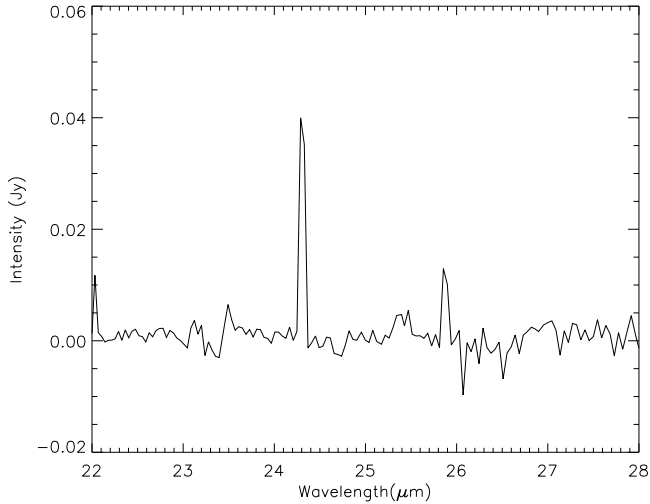}
  \caption{LH infrared spectrum of \TS\ between 22 and 28\,$\mu$m, showing the  [O\,{\sc{iv}}]\,$\lambda$25.89\,$\mu$m  and
    [Ne\,{\sc{v}}]\,$\lambda$24.32\,$\mu$m lines.}
    \label{fig:IR-LH}
\end{figure}  

\TS\ was observed using the Infrared Spectrograph (IRS, \citealt{2004ApJS..154...18H}) on 
board the \emph{ Spitzer Space Telescope} \citep{2004ApJS..154....1W} on 22 April 2006 
(program $\#$20358). The observations used the Short-High 
(SH: 9.9$-$19.6 $\mu$m; R$\sim$600) and Long-High (LH:18.7$-$37.2 $\mu$m; R$\sim$600) 
modules. The aperture of the SH module is $4.7\arcsec\times11.3\arcsec$ and of the
LH one is $11.1\arcsec\times 22.3\arcsec$, so the entire nebular flux was measured. The 
details of the performed observations are shown in Table \ref{tab:log}. 
For LH we used four exposure cycles of 240 s each for on-source and off-source 
observations, while for SH only on-source observations were performed with six exposure 
cycles of 480 s each.  The starting points for our interactive data reduction were the 
co-added 2D flat-fielded BCD (Basic Calibrated DATA)
images (one for each node position; pipeline version 15.3 for 
SH and 17.2 for LH). The rogue pixels were removed using the IRSCLEAN 
tool\footnote{This tool is available from the Spitzer Science Center website: 
http://ssc.spitzer.caltech.edu}, with the {\it aggressive} parameter equal to 0.
Then the data were processed (full extraction, trimming, defringing and averaging over
cycles) into a single spectrum per node position using SMART\footnote{SMART was 
developed by the IRS Team at Cornell University and is available through the Spitzer 
Science Center at Caltech.} \citep{2004PASP..116..975H}. A similar procedure has been applied
for LH off-source observations and the obtained spectra have been subtracted from the
on-source data for the corresponding node position, to cancel out the sky background. The resulting spectrum between  22 and 28\,$\mu$m is shown in Fig. \ref{fig:IR-LH}.
For the high-resolution SH module no background subtraction was done since no sky 
measurements were taken and the SH slit is too small for on-slit background subtraction. 
Finally, the spectra  obtained  for both modules were averaged over two node positions and 
the detected nebular lines measured within SMART. 

The resulting intensities are listed in Table \ref{tab:mid-ir}, together with the 
estimated uncertainties. These uncertainties do not include possible calibration errors. 
It is generally considered that the absolute flux calibration has an accuracy of 20--30\%.
This will be taken into account in the modelling (see Sect. \ref{sec:fluxes-scale}).

Table \ref{tab:mid-ir}  also lists the blue-shifts of the lines. One can see that they are consistent with the optical measurements of Tovmassian et al. (2001). 

\begin{table}
\caption{Observed mid IR line fluxes, in units of $10^{-21}$ W cm$^{-2}$}  
\label{tab:mid-ir} 
\scalebox{0.95}{
\begin{tabular}{lccc}       
\hline 
 ion lambda  &  flux  &  uncertainty  &  rad. velocity \\
    &  $10^{-21}$ W cm$^{-2}$  &  $10^{-21}$ W cm$^{-2}$ & km s$^{-1}$ \\ 
    \hline 
    [O\,{\sc{iv}}]\,$\lambda$25.89\,$\mu$m   &  0.38  &    0.03   &  -180 \cr
    [Ne\,{\sc{v}}]\,$\lambda$24.32\,$\mu$m   &  1.50  &   0.20    &  -140 \cr
    [Ne\,{\sc{v}}]\,$\lambda$14.32\,$\mu$m   &  0.82  &   0.10    &    \\
\hline
\end{tabular}
}
\end{table}

 \subsection{XMM data}
 \label{sec:xmm-data}

\TS\ has been also observed in the X-rays, with XMM. The data acquisition and analysis is presented in \cite{Tovmassian09}.


\section{What else do we know about \TS\ and its exciting star?}
\label{sec:what-else-do}

\subsection{Extinction}
\label{sec:extinction}
\TS\ suffers only little extinction. Using the observed H$\gamma$/H$\beta$ and H$\delta$/H$\beta$ ratios,  \cite{2002A&A...395..929R} had found $E(B-V)\sim 0.3$ mag. However, this estimate was made without considering the underlying stellar absorption in the Balmer lines. Due account for this effect significantly reduces the estimated  $E(B-V)$, as noted by Tovmassian et al. (2004). The extinction can also be estimated by considering the  spectral energy distribution of the stellar core as observed in the far UV by FUSE. Assuming a temperature of 120\,kK for the central star, Tovmassian et al. (2004) obtained a good fit to these observations for $E(B-V)=0.045$ mag, when using non-canonical value for  $R_V$ of 2.3 and the interstellar reddening tables from \cite{1999PASP..111...63F}. Such a low value of $R_V$, as compared to the standard one of 3.1, was considered compatible with the location of \TS\ well outside the galactic disk, since the intervening dust is likely composed of smaller grains than  in the spiral arms. However, we now know that the temperature of the star dominating the UV continuum is much cooler \citep[see Sect. \ref{sec:stellar-core} and][]{Tovmassian09}, implying that a steep reddening law is not needed, after all. In the remainder of the paper, as well as in \cite{Tovmassian09} we use the \cite{1999PASP..111...63F} reddening law parametrized with $R_V = 3.1$, and take  $E(B-V)=0.03$ mag, which satisfactorily account for the observed H and He line ratios as well as the observed continuum. Note that the absence of an absiorption dip at 2200\,\AA\ imposes an upper limit of 0.06 for $E(B-V)$.

\subsection{Expansion velocity}
\label{sec:expansion-velocity}
The expansion velocity of \TS\ has been measured by \cite{2003A&A...410..911R}. This parameter is useful to estimate the expansion cooling in the nebula. It also allows one to have an idea of the nebular  dynamical age. We adopt $v_{\rm{exp}}= 30$\kms.
 
\subsection{The stellar core}
\label{sec:stellar-core}
Our understanding of the stellar core of \TS\ has  evolved considerably since the first paper where it was suggested that \sbs\ is a high excitation planetary nebula (Tovmassian et al. 2001) with a central star having an effective temperature above 100,000\,K. Spectroscopic variations in the course of one single night, reported in Tovmassian et al. (2004), indicated the presence of a  double system with  a compact star. Photometric observations then unambiguously determined a period of 3.92\,h \citep{2005AIPC..804..173N}. Analysis of the light curve indicated that the visible star is likely an elongated ellipsoid irradiated by a source of higher energy. It also  supported the previous conclusion  that the companion must be a (pre-?) white dwarf or a neutron star. Finally, X-ray observations \citep{2007arXiv0709.4016T, 2008AIPC..968...62T} obtained with the XMM-Newton satellite revealed directly the light from the companion, which turns out to be a hot compact star! Thus, as will be shown later, the ``cool'' star is the one visible in the optical and the UV and it provides most of the ionizing photons. But it is the ``hot'' star which gives rise to the high excitation lines observed in the nebula. 
 The best fit to the total spectral energy distribution of the binary core  indicates that the cool component has a temperature $T_{\rm{c}}$  $\sim  60$\,kK while the hot component should have  $T_{\rm{h}}$   $\sim 170$\,kK. However, the determination of the temperature of the hot component is not very accurate.
Note that, in the scenario developed by \cite{Tovmassian09}, the hot component is an \emph{old} white dwarf, which has a 170\,kK temperature not 
because it is still early on its cooling path, but because it was heated by nuclear burning of the 
accreted material on its surface.
For the cool component \cite{Tovmassian09} obtains the following :  $T_{\rm{c}}$ = 58000\, K $\pm$ 3000\, K,  log $g_{\rm{c}} \sim  5.1$. The lower limit on the temperature is the intrinsic temperature of the star, the upper limit corresponds to the zone  heated by irradiation. It is important to note that the cool component is not spherical and has not only an inhomogeneous temperature distribution on its surface but also an uneven 
gravitational acceleration. Its total luminosity is estimated by \cite{Tovmassian09} to be $L_{\rm{c}}$ = 1700\,\lsun with about 30\% uncertainty.
  In the following, for the sake of simplicity, we will consider that the cool star is sufficiently well represented by a stellar model atmosphere with $T_{\rm{c}}=55$\,kK and log $g_{\rm{c}} \sim  5.1$, with a total luminosity of 1700\,\lsun. 
  
  The abundance analysis performed by T.R. on the cool star gives 12 + log He/H = 10.95 and 12 + log C/H = 7.20, with an uncertainty of about 0.3\,dex, and upper limits  12 + log N/H $<$ 6.92 and 12 + log O/H $<$ 6.80.

\subsection{Interstellar absorption of nebular UV lines}
\label{sec:interstellar_abs}

\begin{figure} [h]
   \centering
\includegraphics[width=9cm]{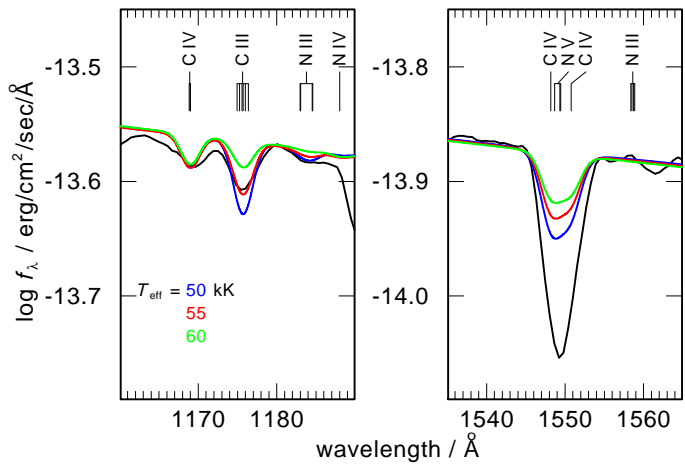}
  \caption{Stellar model atmosphere fitting of carbon lines in the stellar core of \TS. Observation is in black.}
    \label{fig:stellar-C}
\end{figure}  

\begin{figure} [h]
\includegraphics[width=9cm]{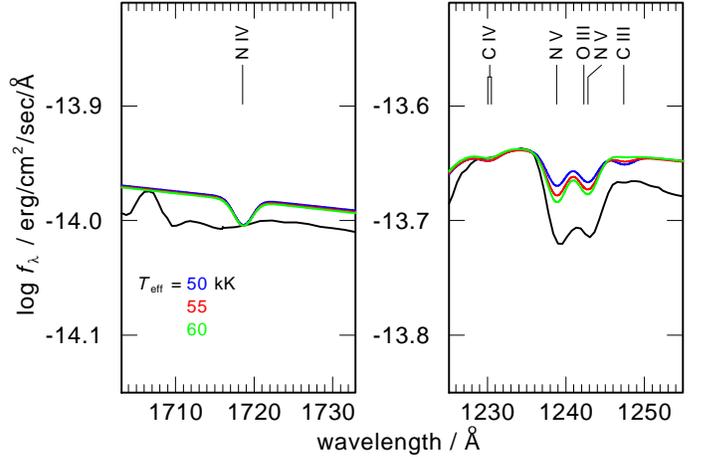}
  \caption{Stellar model atmosphere fitting of nitrogen lines in the stellar core of \TS. Observation is in black.}
    \label{fig:stellar-N}
\end{figure}  

In the course of his stellar atmosphere analysis, T.R. noted that the observed    C\,{\sc{iv}}\,$\lambda$1550\,\AA\ and  N\,{\sc{v}}\,$\lambda$1240\,\AA\ absorption lines were stronger than predicted by his best models. He suggested that these lines are probably affected by interstellar absorption. In that case, the intensities of the C\,{\sc{iv}}\,$\lambda$1550\,\AA\ and  N\,{\sc{v}}\,$\lambda$1240\,\AA\ nebular lines are also affected by absorption.  Since these lines are crucial for the determination of the nebular abundances in \TS, we here explain how we corrected for this effect. 

We use the following notations (all quantities are a function of wavelength):
$F_{\rm{S}}$: flux extracted at the position of the star;
$F_{\rm{N}}$: flux extracted at the adjacent position in the nebula;
$F_*$: real stellar flux;
$F_{\rm{neb}}$: real nebular flux;
$F_{\rm{sky}}$: sky emission and nebular continuum.
The optical depth due to interstellar absorption is denoted $\tau$. 

We have: 
$$F_{\rm{S}} = (F_* + F_{\rm{neb}} + F_{\rm{sky}}) \exp(-\tau)$$ 
and
$$F_{\rm{N}} = (      F_{\rm{neb}} + F_{\rm{sky}}) \exp(-\tau), $$ 
so that, in the spectrum analyzed by R.T., we have: 
$$F_{\rm{S}} - F_{\rm{N}} = F_* \exp(-\tau). $$

Concerning the C\,{\sc{iv}}\,$\lambda$1550 line, reading out from Fig. \ref{fig:stellar-C}, we find $F_* \exp(-\tau) = 8.7 \times 10^{-15}$ (black line in the figure), leading to $\exp(-\tau) =0.75$.
Therefore, if the measured nebular flux  is
$F_{\rm{neb}} \exp(-\tau)=1.28 \times 2.7\times 10^{-14}=3.45\times 10^{-14}$ (last column of Table \ref{tab:HST-int2}), 
the nebular flux after correction for absorption is $F_{\rm{neb}}= 4.65 \times 10^{-14}$ with an uncertainty of about 20\%.

Concerning the  N\,{\sc{v}}\,$\lambda$1240\,\AA\ line, from Fig. \ref{fig:stellar-N}, we find:
$F_* = 2.1 \times 10^{-14}$ (red line in the figure), $F_* \exp(-\tau) = 1.9 \times 10^{-14}$ 
(black line in the figure), leading to $\exp(-\tau) =0.9$.
Therefore, if the measured nebular flux  is
$F_{\rm{neb}} \exp(-\tau)=0.47 \times 2.7\times 10^{-14}=1.27\times 10^{-14}$ (last column of Table \ref{tab:HST-int2}), 
the nebular flux after correction for absorption is $F_{\rm{neb}}= 1.41 \times 10^{-14}$. Here, the uncertainty is larger, since the line is weaker. We adopt 30\%.

\section{Photoionization modelling}
\label{sec:phot-modell}

\subsection{Global strategy}
\label{sec:global-strategy}
The chemical composition of \TS\ can only be determined through photoionization modelling, since we have no direct electron temperature diagnostic.  With the observational data now at hand, we are able to confine the range of possible abundances much better than in previous studies. 
In this paper, we try to make the best use of all the observational constraints. 
The first aspect concerns the morphology. The HST image  (see Fig. \ref{fig:Ha-image}) has an elliptical shape, with two distinct narrow lobes. The fact that the \Ha\ surface brightness distribution shows a hole in the centre  indicates that those lobes are real (possibly due to jets) and not  a thin disk seen edge on. Because the presence of these lobes can affect the ionization structure of the nebula, we choose to carry out the photoionization modelling with a code that allows us to deal with such geometries:  Cloudy\_3D \citep{2006IAUS..234..467M}, based on Ferland's 1D code CLOUDY  \citep{1998PASP..110..761F}. We use version c07.02.01 of CLOUDY and version 594 of Cloudy\_3D.

\begin{figure}
\includegraphics[width=9cm]{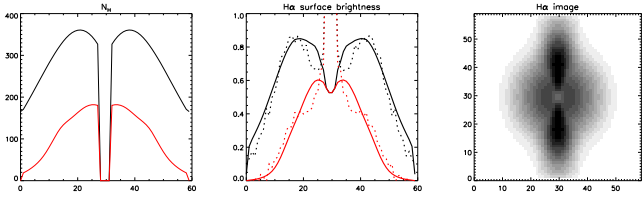}
  \caption{Density structure of the nebula. Left:  The chosen density structure of the model  along the polar axis and along an axis perpendicular to it;  Middle: The resulting \Ha\ surface brightness distribution along the same axes (continuous lines), compared  with the observed distribution (dotted lines); Right:  The theoretical  \Ha\ image.}
 \label{fig:denstruct}
\end{figure}

We assume that the nebula is axisymmetric with its large axis in the plane of the sky, and that the lobes have a circular cross-section. By trial and error, we choose a density structure so as to reproduce the observed  \Ha\ surface brightness distribution.  The chosen density law along the polar axis and along an axis perpendicular to it is shown in the left panel of  Fig.~\ref{fig:denstruct}. The resulting \Ha\ surface brightness distribution along the same axes is shown in the middle panel of this figure (continuous lines), and compared with the observed distribution\footnote {For comparison with the model, we have symmetrised the observed nebular surface brightness.} (dotted lines). The right panel shows the theoretical  \Ha\ image, which can be compared with the observed image shown in  Fig.~\ref{fig:Ha-image}, especially as regards the width of the polar lobes. Note that the density contrast between the lobes and the main body of the nebula is very modest: only  a factor  $\sim$ 2.
The density distribution is parametrized by $n_0$, the value of the density at the center. For each run, we choose $n_0$  so that, within a circle of radius  2.5 \arcsec,  our model returns an \Hb\ flux  of   $2.5\times 10^{-14}$~erg~cm$^{-2}$~s$^{-1}$, which corresponds to the observed extinction-corrected value from \cite{2002A&A...395..929R}\footnote{Note that the models are not ionization bounded.}. The value of $n_0$ is thus dependent on the distance $d$ for which the computations are made. 

The distance $d$, in turn, results from a fitting of the theoretical optical/UV continuum to the observed one (taking into account nebular continuum, aperture effects and reddening). 

For the  stars, we use model atmospheres computed by T.R. with the Tubingen NLTE Model Atmosphere package (TMAP). For the cool component, we use models tailored for our object. For the hot component, in absence of sufficient observational constraints,  we chose among the   complete flux tables for H-Ni  models with halo composition (May 2001) downloaded from http://astro.uni-tuebingen.de/$\sim$rauch. Those models are described in \cite{2003A&A...403..709R}.

\subsection{Putting ultraviolet, optical and infrared fluxes on the same scale}
\label{sec:fluxes-scale}

After a model has been run, the extinction-corrected line intensities are computed for each of the observing slits and compared to the observations.
This is the best way  to deal with aperture corrections, in particular when combining UV and optical, or IR and optical data. Indeed, such a procedure accounts for the ionization structure of the object under study.

Absolute calibration of spectroscopic observations is notoriously difficult. We intercalibrate the UV/optical  data by forcing the measured value of the \Heiiuv/\Heii\  ratio to the one predicted by our photoionization models  in the corresponding slits. The value of $f(\textrm{STIS})$, representing the factor by which the measured  UV fluxes have to be multiplied in order for the \Heiiuv/\Heii\ ratio to be in agreement with the model,  lies between 0.90 and 0.92 in our models. The value of $f(\textrm{STIS})$ is larger for models with larger electron temperature. To allow an easier comparison between models and observations, we fix the value of  $f(\textrm{STIS})$ to 0.91. 

For \textit{Spitzer}-IRS observations, we multiply the observed fluxes by a factor  $f(\textrm{IRS})$ which adjusts the observed values of \Nevir/\Nev\ (after reddening correction) to the one predicted by the photoionization model in the corresponding slit. The values of $f(\textrm{IRS})$ range between 0.87 and 0.95 for the models we considered. It might be judged unreasonable to scale  infrared fluxes using the \Nevir/\Nev\ ratio. However, in the electron temperature domain relevant for \TS, this ratio does not vary very strongly (from $T_\textrm{e}$= 20\,kK to 40\,kK, it decreases by only a factor  2). In any case, this is the only option we have to link together the \textit{Spitzer} line fluxes with the optical ones, since our \textit{Spitzer} data contain no H or He lines. Of course, in the discussion, we bear in mind this difficulty.  To remove the model dependance of the IR fluxes correction, we fix the value of $f(\textrm{IRS})$ to 0.91.

The fact that both $f(\textrm{STIS})$ and $f(\textrm{IRS})$ are found very close to unity is remarkable and means that the flux calibration of the STIS and IRS LH spectra of \TS\ is excellent.

\subsection{Judging a model} 
\label{sec:judging-model}

In order to judge a model, it is convenient to divide the line ratios to be fitted in different categories:
\begin{itemize}
  \item Ratios of hydrogen lines or of helium lines: they probe the reddening law, the stellar underlying absorption, and the recombination line theory.
  \item Ratios of two different ions of the same element, such as \Oivir/\Oiii, N~{\sc v} $\lambda$1240/[N~{\sc iv}] $\lambda$1486, \Nev/\Neiii, and \Nev/[Ne~{\sc iv}] $\lambda$4720. They basically test whether the ionization structure is well reproduced by the model. In this category, we add the \Heii/\Hb\ line ratio, which is more dependent on the ionization level of the nebula than on the abundance of helium. 
  \item Ratios of lines used to determine the chemical composition: \Oivir/\Hb, \Civ/\Hb,   N~{\sc v} $\lambda$1240/\Hb, and \Nev/\Hb. We also consider \Oiii/\Hb\ (although it is redundant with \Oivir/\Hb\  once the ionization structure is reproduced).
  \end{itemize}
Note that, in the case of \TS, the only ratio for direct plasma diagnostics (i.e. electron temperature and density) that is available is \Nevir/\Nev. Unfortunately, the two lines come from measurements in different apertures and with different observing techniques, and, as mentioned above, there is a priori some uncertainty in the relative calibration of the two wavelength domains. However, the fact that we find $f(\textrm{IRS})$ close to unity argues that the electron temperature of the \nev\ emitting zone in our models is not far from the true one.  
  
For all the observables considered (usually line ratios), we compute the value of   
\begin {equation} 
  \kappa(O) = ({\rm log} O_{\rm mod} - {\rm log} O_{\rm obs})/\tau(O),
\end {equation} 
where $O_{\rm mod}$ is the value returned by the model, $O_{\rm obs}$ is the
observed value, and $\tau(O)$ the accepted tolerance in dex for this observable. For each observable, the value of  $\tau(O)$ is chosen \emph{a priori} considering  the observational error bar, including the uncertainty due to reddening,  and the expected ability of our model to reproduce a given observable.  The value of $\tau(O)$ is defined as follows:
\begin {equation} 
   \tau(O) = {\rm log} (1+\Delta  O/O),
\end {equation}
where   $\Delta  O$ is the absolute value of the maximum ``acceptable'' error on the observable. 
 We then judge our models by looking at their outputs presented  in
graphical form (see an example in Fig. ~\ref{fig:losanges}). A model is fully satisfying only if \emph {each} of the
values of $\kappa(O)$ is  found between $-1$ and $+1$, and if the computed line intensities  satisfy the upper limits for undetected lines. Of course, a preliminary condition for a model to be considered is that it returns the correct value of the \Hb\ flux, as explained in Sect. \ref{sec:global-strategy}.

 \begin{figure}
\includegraphics[width=7cm]{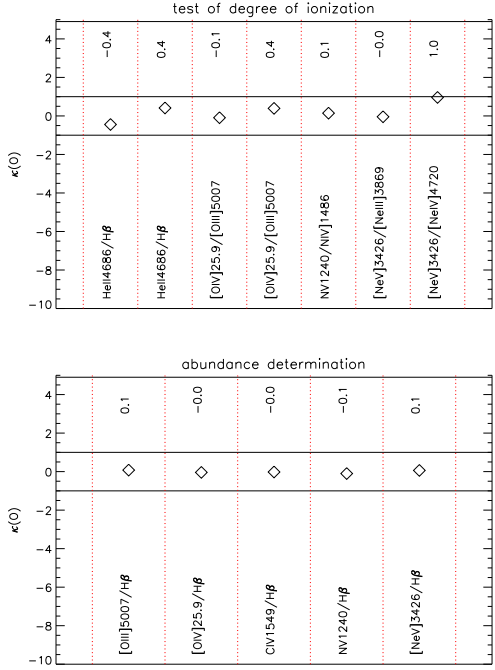}
  \caption{Graphic chart to compare the reference model with the measured line ratios of \TS. The numbers on the top of the panels indicate the values of $\kappa(O)$ for the line ratios listed at the bottom. The line ratios presented in this figure were computed for the following pairs of slits (using the same nomenclature as in Table \ref{tab:models}): (0/0), (4/4), (3/0), (3/4), (1/1), (0/0), and (0/0) for the ratios testing the degree of ionization, and (0/0), (3/4), (1/0), (1/0), and (0/0) for the ratios used to determine the abundances.
   }
    \label{fig:losanges}
\end{figure}

\subsection{The reference model}
\label{sec:reference-model}
Here we present our reference model, R. This is the model for which all the values of $\kappa({\rm O})$ are as close as possible to zero, taking the following characteristics for the cool star:  $T_{\rm{c}}$=55\,kK, log $g_{\rm{c}}$=5.1,  $L_{\rm{c}}$ = 1700\,\lsun, and applying the extinction corrections with $R_V = 3.1$ and  $E(B-V)=0.03$ mag as explained in Sect. \ref{sec:extinction}. The model contains graphite grains (as expected for a carbon rich planetary nebula, which is the case of \TS\ as seen below). The grains have a standard size distribution and  a total dust-to-gas mass ratio of one tenth the standard value. A larger abundance of grains would bring the predicted continuum around  24\,$\mu$m into conflict with the observation. In the following, we explore deviations from the reference model which still account for the observational data.

The reference model has $T_{\rm{h}}$=170\,kK, log $g_{\rm{h}}$=6.7, and a total luminosity $L_{\rm{h}}=$2564\,\lsun\footnote{The mass of the hot stellar component in the models depends on the value assumed for the gravity, which is not well constrained. For the photoionization modelling, what really matters is $T_{\rm{h}}$ and $L_{\rm{h}}$. };  $n_0$=181\,cm$^{-3}$ and the following abundances, in units of 12 + log (X/H): He=10.95, C=7.84, N=7.15, O=6.82, Ne=6.83, S=5.65, Ar=4.70.  

The models and observations to which they are compared are presented in Table \ref{tab:models}. Column 1 of this table lists the line identifications, Column 2 characterizes the observation using the following nomenclature: 0 for CFHT 2003, 1 for STIS, 2 for FUSE, 3 for \emph{Spitzer}, 4 for SDSS, 5 for CFHT 2001. For lines which belong to a wavelength range that was not observed the number 6 is attributed.   Column 3 lists the  observed reddening-corrected line intensities (or their upper limits), in units of  \Hb=100 in the corresponding apertures. Column 4 lists the acceptable relative error $\Delta$O/O used to compute $\kappa(O)$. In the case of HST, FUSE and \emph{Spitzer} data, we estimate the value of \Hb\  in the relevant aperture, basing on our models (since, as explained in Sect. \ref{sec:phot-modell}, they deliver a smoothed version of the observed surface brightness distribution). The top rows of column 5 of the table list the characteristics of the reference model. The predicted line intensities in the relevant aperture  are given in the following rows, in units of  \Hb=100 in the same aperture. 
For easier analysis,  the next rows list a few important line ratios, the intensity of each line being measured through the  aperture corresponding to the observation. In order to shorten the table, we do not list the lines for which the predictions from all our models give values smaller than 0.001 of \Hb\ (we note that such is the case for all the recombination lines of elements C, N, O). We list only the strongest H and He lines (we checked that the weaker \hi\ and \heii\ lines always give $|\kappa(O)| < 1-1.5$ in our models, implying that the correction for stellar absorption and reddening is satisfactory).

 The graphical representation of the line ratios predicted by  model R and used to estimate the chemical composition of \TS\ is shown in Fig. ~\ref{fig:losanges}.  This is the kind of chart that was used in practise when judging the models that were run. A ``best model'' is one for which all the diamonds fall as close as possible to ordinate 0.  In any case, an acceptable model should have all line ratios represented by a diamond between the two horizontal lines, which represent a one-sigma deviation from the observed value. In addition, acceptable models should not return line intensities above the upper limits allowed by the observations. 
 
 Figure \ref{fig:images} shows the monochromatic images of the reference model in various emission lines. It reveals a few interesting features of the model:  some lines, such as  \Oiii\ and  \Neiii\  arise mainly in the lobes. Other lines, such as \Civ,  N~{\sc v} $\lambda$1240, and \Nev\ line come from the entire nebula (in agreement with what Fig. \ref{fig:SBS_3433} suggests for \Nev), while  O~{\sc vi} $\lambda$1032 and [Ne~{\sc vi}] $\lambda$7.6$\,\mu$m (the latter not in the observed wavelength range) come from the innermost regions. We can also see that the two  C~{\sc iii} lines, C~{\sc iii}] $\lambda$1909 and   C~{\sc iii} $\lambda$977, although produced by the same ion, come from different regions: C~{\sc iii} $\lambda$977 has an important component coming from the central main body  (see Fig.   \ref{fig:images}), where the very high electron temperature allows for its excitation even if C$^{++}$ is not very abundant there.

%

\begin{table*} [hb!]
\caption{Photoionization models versus observations. Line intensities are in units of H$\beta$=100 in the corresponding aperture} 
\begin{center}
\scalebox{0.85}{
\begin{tabular}{rr@{\hspace{5mm}}r@{\hspace{-2mm}}r@{\hspace{1mm}}r@{\hspace{10mm}}@{\extracolsep{5mm}}rrrrr}      
\hline
\input{table5-11sept-final.txt}
\hline		
\multicolumn{10}{l}{$^a$ Labels for the observing slits: 0: CFHT 2003; 1: STIS; 2: FUSE; 3: \emph{Spitzer}; 4: SDSS; 5: CFHT 2001;	}\\														
\multicolumn{10}{l}																
{$^b$ Asplund, Grevesse	\& Sauval (2005) ; $^c$ abundances in units	of 12 + log X/	H ;	$^d$ in units of 10$^{-14}$ \ergcmsq.	}\\	

\hline 
\end{tabular}}
\label{tab:models} 
\end{center}
\end{table*}

%
 \normalsize

\begin{figure}
\includegraphics[width=9cm]{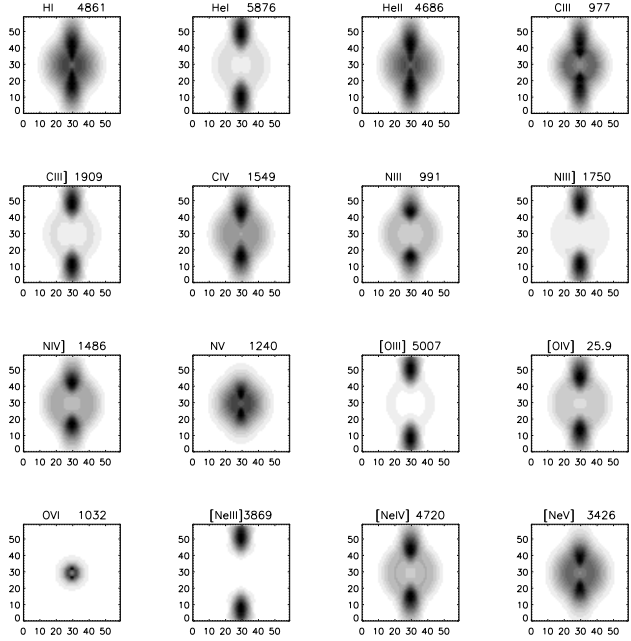}
\caption{Monochromatic images of the reference model in various lines (the values of the wavelengths are in \AA\ if they are larger than 900, and in $\mu$m otherwise). The $x$ and $y$ values are the coordinates in pixel units of the  models. }
\label{fig:images}
\end{figure}

Figure \ref{fig:sed} compares the observed energy distribution from \TS\ with the computed one.  The top panel shows the reddening-corrected flux computed for model R, in erg cm$^{-2}$ s$^{-1}$ \AA$^{-1}$, in the wavelength  range 900\,\AA$-$40\,$\mu$m. The observations  are superimposed in various colours, as indicated in the caption. One can see that the model reproduces the observed spectral energy distribution quite well (except for the IRS SH observations which could not be corrected for sky emission, as explained in Sect. \ref{sec:mid-infr-spectr}).  The bottom panel shows the energy distribution in the soft X-ray range: the blue curve is the hot star, the green triangles are the XMM observations. The flux from the star has been corrected for the nebular absorption (computed by CLOUDY) and for the interstellar absorption, taking a hydrogen column density of $1.6\times 10^{20}$ cm$^{-2}$. For each computed model  we checked that the ionizing flux does not violate the observed stellar emission up to 200\,eV. At higher energies, the observed emission may have another origin than the stars we consider, but it does not affect our model fitting, since we have no relevant observational constraints (the ion with highest ionization potential observed is {Ne$^{4+}$}, which has an ionization potential of 97.1\,eV).    

Figure ~\ref{fig:sed2} compares the energy distributions of the two stars considered in the modelling: the ``cool'' star is in red, the ``hot'' one is in blue. The sum of the two is in black. As mentioned in Sect. \ref{sec:stellar-core}, the cool star dominates in the optical range (a few eV), and until about 20 eV, but it is the hot star which provides the photons with energies above $40-50$\,eV. So that it is the cool star which provides most of the H ionizing photons, but it is the hot star which provides the photons responsible for the presence of the \heii, N~{\sc v}, [O~{\sc iv}], [Ne~{\sc iv}], \nev\ and [Ar~{\sc v}] lines. This is a very uncommon situation, perhaps a unique case among planetary nebulae: \TS\ has two ionizing stars! This explains why our previous attempts to model the object were facing the difficulty that the nebula needed plenty of photons of energies above 54.4 eV,  while the Balmer absorption lines in the stellar continuum indicated a moderate temperature. 

From Figs. \ref{fig:losanges},  \ref{fig:sed}, and Table \ref{tab:models},  one can see that our reference model fits all the observational constraints very well. The only exception is that of the C~{\sc iii}] $\lambda$1909, which is slightly above the upper limit we gave to the STIS observation. However, we consider this result to be still acceptable, since upper limit fluxes for unobserved lines are difficult to estimate accurately. The abundances of C, N, O, and Ne in the reference model are, respectively, 1/3.5, 1/4.2, 1/70, and   1/11, and, for S and Ar $<$ 1/30 of the Solar values given by  \cite{2005ASPC..336...25A}.

  \begin{figure}
\includegraphics[width=9cm]{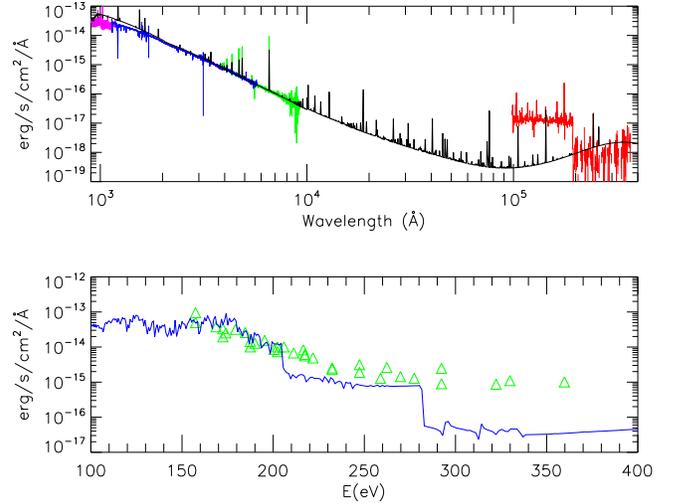}
  \caption{Comparison of the reddened spectrum of the reference model with observations. \textbf{Top:} From the UV to the IR. The reddened model is in black. The colour code for the observations is as follows. Magenta: FUSE; blue: HST; green: SDSS ; red: Spitzer (the SH observations could not be sky-corrected).   \textbf{Bottom:} X-ray domain. The model (with extinction applied)  is in blue. The XMM observations are represented by triangles. }
    \label{fig:sed}
\end{figure}

  \begin{figure}[h!]
\includegraphics[width=9cm]{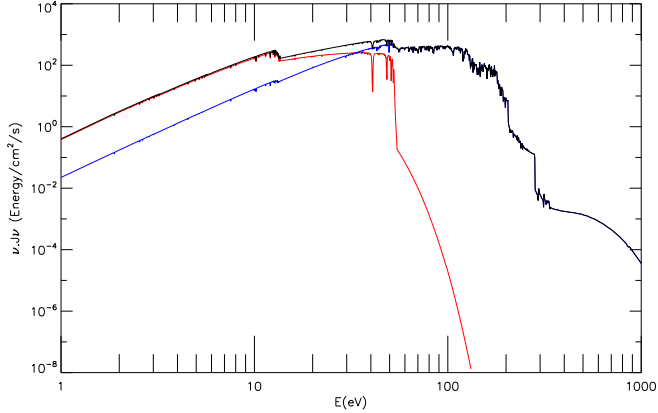}
  \caption{\textbf Spectral energy distribution (normalized to arbitrary units) of the radiation from the ionizing stars in the reference model. Red: the ``cold'' star; blue: the ``hot'' star; black: the sum of the two.  }
    \label{fig:sed2}
\end{figure}

\section{The chemical composition of \TS}
\label{sec:chem-comp-png}

\subsection{Range of abundances for the reference model}
\label{sec:range-abund-refer}

We now investigate the error bars on abundances that are due only to the uncertainties in the observed line intensities. Since C, N, O, and Ne contribute negligibly to the energy budget, it is straightforward to estimate from model R the minimum and maximum abundances corresponding to the minimum and maximum values of \Oivir/\Hb, \Civ/\Hb,   N~{\sc v} $\lambda$1240/\Hb, and \Nev/\Hb\  without changing the ionization structure of the nebula. However, one has also to consider the  error bars on line ratios that constrain the ionization structure:  \Oivir/\Oiii, N~{\sc v} $\lambda$1240/[N~{\sc iv}] $\lambda$1486, \Nev/\Neiii, and \Nev/[Ne~{\sc iv}] $\lambda$4720. The minimum values of the C, N, O, and Ne abundances in  \TS\ are obtained by a model with the lowest ionization compatible with the observations and the lower limits of \Oivir/\Hb, \Civ/\Hb,   N~{\sc v} $\lambda$1240/\Hb, and \Nev/\Hb.
Such a model, Mi, is reported in column 6 of Table \ref{tab:models}. It is derived from the reference model R by lowering the values of $L_{\rm{h}}$, and decreasing the values of the C, N, O, and Ne abundances. With similar considerations, one can construct a model Ma, which will give the maximum C, N, O, and Ne abundances. This model, with a higher $L_{\rm{h}}$ and same $T_{\rm{h}}$ is listed in column 7 of table~\ref{tab:models}.

The resulting limits on the abundances of C, N, O, and Ne in the gaseous phase\footnote{The contribution of grains to the abundance of carbon is discussed in Sect. \ref{sec:dust}.} of \TS\ are thus:
\begin{description}
\item[ ] 7.64 $<$	12 + log C/H$<$	8.05
\item[ ] 7.00	 $<	$ 12 + log N/H $<$ 7.32
\item[ ] 6.63	$<$ 12 + log O/H $<$	7.13
\item[ ] 6.76	 $<$ 12 + log Ne/H $<$	6.90
\end{description}

The limits on the C/O, N/O and Ne/O ratios are obtained by considering tailored  models reproducing the extreme values of the observed intensities of C, N and Ne lines. They are:  
\begin{description}
\item[ ] $ 0.83 <$ log C/O $< 1.21$
\item[ ] $ 0.12 <$ log N/O $< 0.54$
\item[ ] $-0.16 <$ log Ne/O $< 0.17$
\item[ ] 
\end{description}

To derive the lower limit on He/H, one must consider the model with the lowest ionization compatible with  the observed \Oivir/\Oiii, N~{\sc v} $\lambda$1240/[N~{\sc iv}] $\lambda$1486, \Nev/\Neiii, and \Nev/[Ne~{\sc iv}] $\lambda$4720, and the lowest \Heii/\Hb. The upper limit on He/H is obtained with similar arguments. The corresponding models, HeMi and HeMa, respectively, are listed in columns 8 and 9 of Table \ref{tab:models}. The resulting limits for He/H obtained in this way are 0.095 and 0.081. In other words, the precision on the He/H abundance is not very good, despite the fact that we have been fitting the \Heii/\Hb\ ratio
within 4\%  (the formal uncertainty in this ratio is  2\% for the 2003 CFHT data, but comparison with SDSS data led us to adopt a higher value for the tolerance). With a more accurate upper limit on \Hei/\Hb\, we could reduce the error bar on the helium abundance (as a matter of fact, model HeMi slightly violates the present upper limit on \Hei/\Hb). But the uncertainty in He/H will remain larger than the uncertainty in the \Heii/\Hb\  ratio, mainly because the electron temperature gradient in this nebula is large and the ratio of emissivities of \Heii\ and \Hb\ slightly varies with temperature.

\subsection{Additional sources of abundance uncertainties}
\label{sec:addit-uncert-abund}

In this section, we discuss how reasonable variations of the parameters that were so far fixed in the modelling procedure affect the derived abundances.   We also discuss some more general problems that may have an influence on the estimated chemical composition of \TS. To save space, the models that were constructed to discuss these additional uncertainties are not listed in the paper. We will only mention their impact on the derived abundances. Note that  all of those additional models have the required angular size and total H$\beta$ flux, and their abundances have been chosen so as to fit  the observed emission line ratios.

\subsubsection{The effect of changing the description of the stars}
\label{sec:cool-star}
So far, we have kept the parameters of the cool star fixed. Even if they are rather well determined, as explained in Sect. \ref{sec:stellar-core} and \cite{Tovmassian09}, it is important to see the effect that a change in those parameters implies on the derived chemical composition of the nebula. It turns out that an increase of 5\,kK in $T_{\rm{c}}$ induces a decrease in the C, N, O and Ne abundances by 0.05--0.08\,dex. A change in   $g_{\rm{c}}$ by 0.1\,dex leaves the abundances of the fitted nebular model unchanged. 

We then explored the effects of changing the parameters of the hot star. By increasing  the temperature of the hot star by 10\,kK,  (implying a slight decrease of its luminosity in order to fit the observations)  one decreases the C, N, O, Ne abundances of the fitted nebular model by about 0.2\,dex. 

We also explored the effect of changing the model atmosphere of the hot star. One extreme case is to consider a model atmosphere composed only of H and He, instead of the Galactic halo chemical composition. Consequently, there are no absorption edges in the atmosphere above 54\,eV. The model which fits the observations presents an intense [Ne~{\sc vi}] $\lambda$7.6$\,\mu$m emission (unfortunately outside the wavelength range of our IRS observations of \TS). Its Ne abundance is higher by 0.15\,dex than that of the reference model, while the abundances of the remaining elements are almost unchanged. 

\subsubsection{Dust issues}
\label{sec:dust}

Concerning extinction and reddening issues, changing $E(B-V)$ and $R_V$ within limits compatible with the observed Balmer decrement and the observed stellar energy distribution does not change the abundances derived for \TS\ significantly. 

Our reference model has a dust-to-gas mass ratio of $10^{-1}$ times the canonical value,  with the canonical grain size distribution as stated in Sect. \ref{sec:reference-model}. The chemical composition of the grains --pure graphite-- is dictated by the fact that the object is undoubtedly carbon-rich, as seen in Sects. \ref{sec:stellar-core} and \ref{sec:range-abund-refer}. The total abundance of grains in the reference model is chosen so that the predicted infrared flux arising from the heated grains does not exceed the observed IRS LH flux and that it produces no significant   dip at 2200\,\AA\ since this is not observed. The total amount of carbon locked in grains in the reference model is 0.4 times that of the abundance of carbon in the gas phase.  This means that the total abundance of carbon in the nebula (gas  plus grains) is larger  by about 0.15\,dex than given in Sect. \ref{sec:range-abund-refer}. 

\subsubsection{The role of morphology}
\label{sec:morphology}

While constructing our reference model (and all the models described before), we have chosen a geometry that reproduces the observed \Ha\ surface brightness, including the lobes. It is interesting to experiment a simpler model without any lobes, in which the averaged surface brightness is the same as in the reference model. The abundances in such a model differ insignificantly from those of the reference model. We have to confess that we were somewhat surprised by this result, since as shown in Fig. \ref{fig:images}, the emission in such lines as \Oiii\ or \Neiii\ traces the lobes very distinctly. On the other hand, one has to remember that the density contrast between the lobes and the ambient medium is only a factor 2, as seen in Fig. \ref{fig:denstruct}.

\subsection{Caveats}
\label{sec:caveats}

\subsubsection{The problem of H$\alpha$}
\label{sec:Ha}

One of the intriguing problems in the observations of \TS\ is the behaviour of the \Ha\ line. As seen in Table \ref{tab:optical-lines}, the observed \Ha/\Hb\ ratio varies among data taken during different runs and at different telescopes. Since the ratios of all the remaining hydrogen lines look normal, within the error bars, we are inclined to think that this \Ha\ problem has no influence on the derived chemical composition. Nevertheless, we feel it important to try to understand the reason for the observed values of \Ha/\Hb.

In the present study, we have done the computations with the full treatment of hydrogen as offered by CLOUDY (this, and not case B,  is actually the default option in CLOUDY). Under the  physical conditions in this nebula, one indeed does not expect the Balmer lines to be emitted under case B, not even with the added effect of collisional excitation. The ionization parameter of the emitting regions is high and the nebula is optically thin, making it a good candidate  for case C as described by \cite{1938ApJ....88...52B} and  reconsidered by \cite{1999PASP..111.1524F}. In such a case, absorption of Lyman photons from the star contributes to the emission of the Balmer lines, and the  Balmer decrement depends on the number of respective Lyman line photons in the star. However, we are far from reproducing  the  \Ha/\Hb\ ratios observed in the various slits. Of course, the computed Balmer decrement strongly depends on the fluxes at the wavelengths of the H Lyman lines in the model atmosphere used. But the differences in the  \Ha/\Hb\ ratios in the different observing runs make it doubtful  that simple stellar fluorescence can explain the observations.

The reference model  predicts a ratio of about 2.81.  
The differences in the  observed \Ha/\Hb\ ratios  cannot have a nebular origin since the associated time scales are far too long.

Water vapour absorption near \Ha\ is far too weak to explain the variations\footnote{http://www.astrossp.unam.mx/sitio/abs\_telurica\_english.htm}. 
Now that the nature of the binary central star is better known, we can also discard the possibility that much of the \Ha\ emission comes from an accretion disk.  Active mass transfer in the system has ceased and, even if there is a stellar wind or weak remnant of accretion disk around the hot component, it cannot have a big influence on emission lines, since we detect fairly symmetric underlying absorption lines from the cool component at all orbital phases.  These symmetric lines also imply that extra emission from the irradiated face of the cool component  does not contribute significant \Ha\ emission.

The remaining option is atmospheric refraction \citep{1982PASP...94..715F}, since the slit was not oriented at the parallactic angle for many (though not all) of the spectra with \Ha/\Hb\ ratios differing significantly from 2.81 (rather, usually east-west).  What is odd,
a priori,
if atmospheric refraction is responsible, is that the lines from H$\delta$ to \Hb\ are observed with constant intensity ratios.  
Simulations in which we convolve the quantum efficiency of the slit camera used at SPM\footnote{http://www.astrossp.unam.mx//Instruments/bchivens/camrend/manual-english.pdf} with the object's very blue spectrum indicate that the effective wavelength is
 between 4000\AA\ and 4500\AA.  Thus, the effective wavelength, which is what is used to centre the object in the slit, is between the blue lines, so atmospheric refraction has very little effect upon them.  As a result, \Ha\ should be the only optical line that may be significantly affected by atmospheric refraction.  Also, compared to the usual assumptions, the wavelength baseline over which atmospheric refraction operates is unusually large in this case, of order of 2000\,\AA\ or more.  Tests using the SPM4 dataset \citep{2002A&A...395..929R}, in which this issue can be studied in greatest detail, clearly implicate the effect of atmospheric refraction since the spectral shape of the central star's continuum varies as a function of the difference between the slit position angle and the parallactic angle.   Therefore, we are inclined to attribute the variations observed in the \Ha/\Hb\ ratio to atmospheric refraction.

\subsubsection{Atomic data}
\label{sec:atomic-data}

 As noted by  \cite{2005A&A...430..187P}, the atomic data on which  photoionization models are built are not of perfect accuracy. All the models we have computed rely on CLOUDY c07.02.01. It is not excluded that future advances in atomic physics, especially in the calculation  of recombination coefficients for highly ionized species, might affect the computed ionization structure. However, the fact that we now have observational data (or stringent upper limits) on several ions of each of the elements C, N, O and Ne in \TS\  makes us confident in the robustness of the chemical composition that we have derived. The relatively large error bars we are obtaining on the abundances (principally due to the lack of a direct measure of the electron temperature in the nebula) imply that the  uncertainties in atomic data, including the collision strengths of the lines used for the diagnostics, should be negligible in the total error budget.  

\subsubsection{Dynamical effects}
\label{sec:dyn-effects}
\cite{2005AIPC..804..269S} have drawn attention to the possible importance of dynamical effects in the thermal balance of nebulae. They make the point that the role of dynamical expansion in the cooling budget increases as the metallicity decreases. We have therefore included the effect of expansion cooling in CLOUDY, by introducing a wind cooling contribution in the routine \verb"cool_eval.cpp": 
\begin{verbatim}
dynamics.dDensityDT = (float)(2.*fudge(0));
CoolHeavy.expans = 
dense.pden*phycon.te*BOLTZMANN*dynamics.dDensityDT;
\end{verbatim}
with the user defined parameter ``fudge'' being related to the expansion velocity and the outer radius of the nebula by \verb"fudge"= $v_{\rm{exp}}$/$R_{\rm out}$.

All the models presented above have been computed with an expansion velocity of  30\kms, corresponding to the observed value (see Sect. \ref{sec:expansion-velocity}). We have tried other values for  $v_{\rm{exp}}$ in the equation above, but noted no significant changes in the output between  0 and 200\kms, the extreme values we tried.
This result is at variance with the finding by  \cite{2005AIPC..804..269S} that expansion cooling significantly reduces the temperature with respect to a fully static model of same density structure.

In our models, the dominant cooling process is collisional excitation of H Ly$\alpha$, and, at the ionization level predicted by the model, it is clear that expansion cooling must be negligible, unless the velocity of the jet is of the order of  1000\kms.

Could it be that the lower temperature found by \cite{2005AIPC..804..269S} in fully dynamical models with respect to hydrostatic ones, which they attribute to expansion cooling, is actually the result of some other process? The only idea that comes to mind is departure from ionization equilibrium. For an average temperature of 30\,kK,  and  an average density of about 200\cmcub, the recombination time for hydrogen is of about $10^3$\,yr.
From the apparent size, expansion velocity and distance to TS01, one can estimate an expansion time of  $\sim$ $7\times 10^3$\,yr. Therefore, the nebula should not be far from ionization equilibrium. 
On the other hand, the dynamical model shown in \cite{2005AIPC..804..269S} was  for a 0.595\msun\ star with an effective temperature of 100\,kK,  corresponding to an evolution time of $\sim$ $5\times 10^3$\,yr. The average density of the nebula in their simulation is then about 100\cmcub. In such a situation, the nebula is  farther from ionization equilibrium. Since their star is in a phase where the number of ionizing photons increases with time, the ionization level of the dynamical model should be smaller than that of the corresponding static model. Hence, Lyman alpha cooling should be more important, and the electron temperature lower than in the hydrostatic model, which is indeed what their dynamical model gives. In \TS\ the dynamical effects on the ionization and temperature of the nebula should be much smaller than in the case computed by Sch\"onberner et al., if noticeable at all. In their model, the temperature drop due to dynamical effects is of about 10\,kK. Given the argumentation above, we consider that any dynamical effect on the electron temperature in \TS\ would be of 2--3\,kK at most, with respect to the temperature we compute in our model. As an experiment, we computed a model where we use the CLOUDY parameter \verb"cextra" with a value of $10^{-20.3}$ erg cm$^3$ s$^{-1}$ to simulate an  extra cooling factor that reduces the average electron temperature by about 3\,kK with respect to the reference model.  We then adjust the abundances so as to reproduce the observed line ratios. We find that the abundances in this model are not very different from those of model R. In particular, the abundance of O is not changed. The reason is that the model must reproduce the \Nev/\Neiii\ ratio, which is nearly temperature-independent, and that ratios  used to constrain the oxygen abundance ([O~{\sc iv}] $\lambda$25.9$\mu$m/\Hb\ and \Oiii/\Hb) are not very sensitive to the temperature above 30\,kK.

\subsection{Wrapping up}
\label{sec:wrap-ab}

In summary, considering all the possible sources of uncertainties, and adding in quadrature the various independent errors, we find that the elemental abundances in the gas phase  of \TS\ are as listed in 
Table \ref{tab:abund}.

\begin{table} [hb!]
\caption{Nebular abundances of TS 01, in various units}  
\label{tab:abund} 
\scalebox{0.95}{                    
\begin{tabular}{lrlcc}       
\hline\hline 
	&	12+log X/H	&	uncertainty	&	X/H	&	mass fraction	 \\
\hline 
He	&	10.95	&	$\pm$ 0.04	&	$8.91\times 10^{-2}$	&	$2.63\times 10^{-1}$	 \\
C	&	7.84	&	$\pm$ 0.30	&	$6.92\times 10^{-5}$	&	$6.11\times 10^{-4}$	 \\
N	&	7.15	&	$\pm$ 0.25	&	$1.41\times 10^{-5}$	&	$1.46\times 10^{-4}$	 \\
O	&	6.82	&	$\pm$ 0.33	&	$6.61\times 10^{-6}$	&	$7.79\times 10^{-5}$	 \\
Ne	&	6.83	&	$\pm$ 0.30	&	$6.76\times 10^{-6}$	&	$9.96\times 10^{-5}$	 \\
S	&	$<$ 5.5  	&	 	       &	$<$ $3.16\times 10^{-7}$	&	$<$ $7.45\times 10^{-6}$	 \\
Ar	&	$<$ 4.5 	&	 	       &	$<$ $3.16\times 10^{-8}$	&	$<$ $8.38\times 10^{-7}$	 \\
\hline                               
\end{tabular}
}
\end{table}

An additional amount of carbon, about 40\% of the total elemental abundance, is locked up in dust grains. Allowance for this component  raise the carbon abundance in the nebula to 12 + log C/H = 8.00$\pm$0.3. 

The error bars on the derived abundances may seem  large, when compared to the typical error bars in other PNe. However, one must remember that the analysis of \TS\ is much more difficult, due to the absence of direct temperature diagnostics and to the weakness of the lines from metals. 

The abundances derived for the nebula are consistent with those derived by T. R. for the atmosphere of the cool star, see Sect. \ref{sec:stellar-core}, except for carbon whose abundance is larger by 0.8\,dex in the nebula. This agreement is remarkable, given the difficulty of the analysis. Whether the discrepancy between the carbon stellar and nebular abundances is real should be examined in more detail.

\section{Understanding the observed chemical composition}
\label{sec:png-understanding} 
 
\subsection{\TS\ compared to other Galactic halo PNe}
\label{sec:png-compared-other}

\begin{figure*}
\includegraphics[width=18cm]{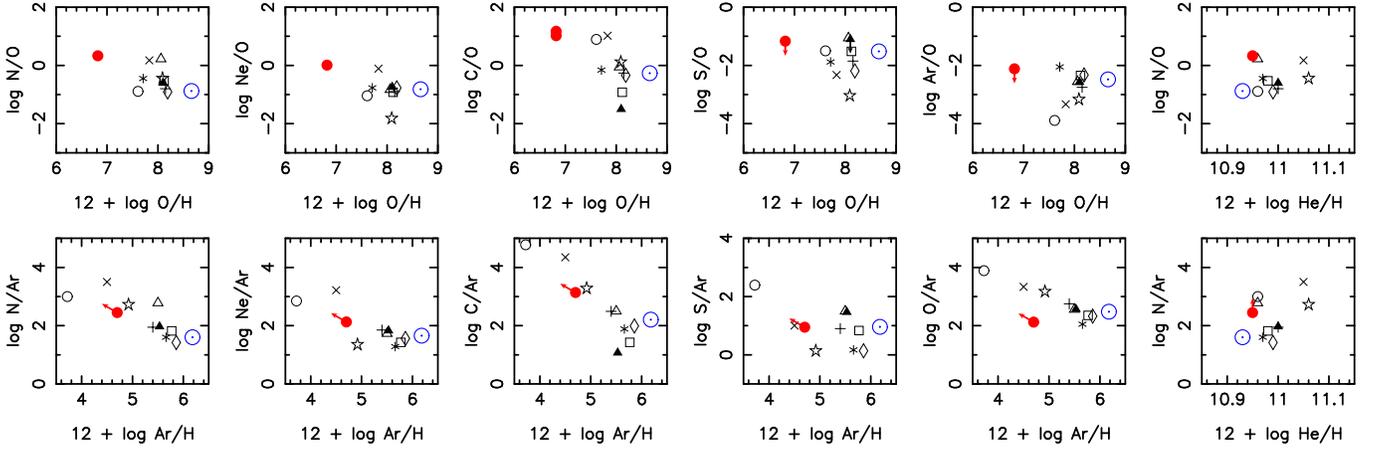} 
  \caption{Comparison of \TS\ (filled red circle) with other Galactic PNe in abundance ratios diagrams. The different black symbols represent: K648 (circle), DdDm-1 (square), PN006-41.9 (triangle), NGC 2242 (diamond), NGC 4361 (plus sign), M2-29 (asterisk), H4-1 (star), PN 243.8-37.1 (filled triangle), and BoBn-1, which actually belongs to the Sagittarius dwarf spheroidal galaxy ($\times$ sign). The Sun is represented by the Solar symbol.}
    \label{fig:HALOPN}
\end{figure*}

Figure ~\ref{fig:HALOPN} shows the abundance pattern of \TS\ with respect to those of other PNe located in the Galactic halo as  derived by \cite{1997MNRAS.284..465H}. The panels in the top rows  of this figure show the elemental abundances with respect to oxygen, as a function of O/H, a common way to display abundances in PNe. As has been noted before, PNe belonging to the Galactic halo display a large dispersion in their abundances relative to oxygen\footnote{The dispersion in S/O, however, is likely significantly affected by important errors in the sulfur abundances.}. In this respect, \TS\ is not an exception. But it is extreme in its value of O/H which is significantly lower than  in other PNe of the Galactic halo. On the other hand, the C/O, N/O, Ne/O ratios are similar to the highest ones found in  those objects. In the panels of the  bottom row, the abundance ratios are computed with respect to argon and displayed as a function of 12 + log Ar/H. The reason for doing this is that the abundance of Ar is not expected to be modified with respect to the initial chemical composition out of which the progenitor star was formed. Of course, the determination of the argon abundance is much less accurate than that of oxygen. In the case of \TS, we even have only an upper limit. Yet, the bottom row of Fig. ~\ref{fig:HALOPN} confirms the impression that, regarding the chemical composition, \TS\ is different from the remaining halo PNe especially because of its remarkable low oxygen abundance. It may have the lowest Ar/H as well, but we do not know, since we have only an upper limit on its abundance. 

As can be seen from Fig.~\ref{fig:HALOPN}, the progenitors of many of the PNe belonging to the Galactic halo underwent considerable nuclear processing, which affected not only the helium, carbon and nitrogen abundances in the nebulae, but also the oxygen one. In such a situation, the metallicity of the progenitor --generally identified with the oxygen abundance in the nebula-- cannot be easily determined. If one uses the sum of the mass-fraction abundances of C, N, O, Ne, S, and Ar as a proxy for the upper limit of the metallicity, we find that, in Solar units, \TS\ has a metallicity of at most 1/12 of Solar, well below the upper limits for the remaining halo PNe. If we take the argon abundance as a proxy for the metallicity, we find that the metallicity of \TS\ is less than 1/30 of Solar. The two other halo PNe with very low metallicity, using this criterion,  are BoBn-1 (1/20 of Solar) and K 648, which holds the record from this point of view (1/300 of Solar).

\subsection{What we know of the progenitor of \TS}
\label{sec:progenitor}

Let us first summarize the main features of a possible evolution of the stellar core of \TS\ as explained in Tovmassian et al. (2007, 2008, and 2009). The total mass of the system is close to 1.4\,\msun.\ The mass of the cool component is
(0.5--0.6)\,\msun. Then the mass of the hot component must be (0.8--0.9)\,\msun. For Z$\simeq$ 0.001 this implies that its progenitor had a mass of
$\lesssim 3$\,\msun\  \citep{hpt00}. Star formation in the halo ceased at least 10\,Gyr ago, see e.g. \cite{2009ApJ...694.1498M}.
Since the lifetime of a (2 -- 3) \msun\ star with Z $\simeq$ 0.001 is $\lesssim 300$ Myr,  the first Roche lobe overflow occurred at least 10\,Gyr ago. It resulted in a common envelope with inefficient accretion, hardly more than
$\sim$ 0.01\,\msun\ \citep{1995ApJ...447..656Y}. The initially more massive star of the system  turned into the currently hot  component. Since we observe a PN \emph{now}, the companion of the hot component must have completed its evolution in 10\,Gyr. The least massive star able to do this had a mass of about
 0.9\,\msun\ after the first common envelope stage.Ê
When this  low mass star evolved  off the main sequence,
a symbiotic system formed. Mass transfer during this stage was negligible as well. The symbiotic stage ended when the low mass star filled its Roche lobe on the AGB. Again,
common envelope formed, reducing the separation of the components and diminishing the orbital period to 3.9 hr. The nebula we observe now is the remnant of this second common envelope.

\subsection{Can standard single AGB models explain the observed abundance pattern?}
\label{sec:yields-standard}

We consulted the available
 yields from recent standard
 evolution models for single stars at low metallicities  
 (see \citealp{KarakasLattanzio2003}
 and \citealp{CristalloStraniero2009}).
The corresponding nucleosynthetic predictions for low-mass AGBs show
the signature of recurring third dredge-up  episodes that increase
the surface abundances of C as well as (but to a lower extent) that of O and N. This is at odds with the abundance patterns determined in \TS, which exhibits an extreme  O depletion.
As a matter of fact,
only massive AGB models ($M_{\rm ini}\ge4$\msun)
predict O depletion as a result of hot bottom burning (HBB)
during the thermal pulse AGB phase (TP-AGB). Besides being in conflict
with the binary analysis recalled above, these massive AGB models are
facing other difficulties. First,
they predict that carbon also should be depleted by the CNO-cycle
in the envelope.
Second, such massive AGB stars are predicted to
experience deep second dredge-up 
that increases the He abundance
well above the observed value (number ratio of ${\rm He/H}\sim0.093$, thus
  ${\rm Y}\sim 0.26$ in mass fraction).

\subsection{Towards an explanation of the chemical composition of \TS}
\label{sec:path}

\begin{figure}
\includegraphics[width=9cm]{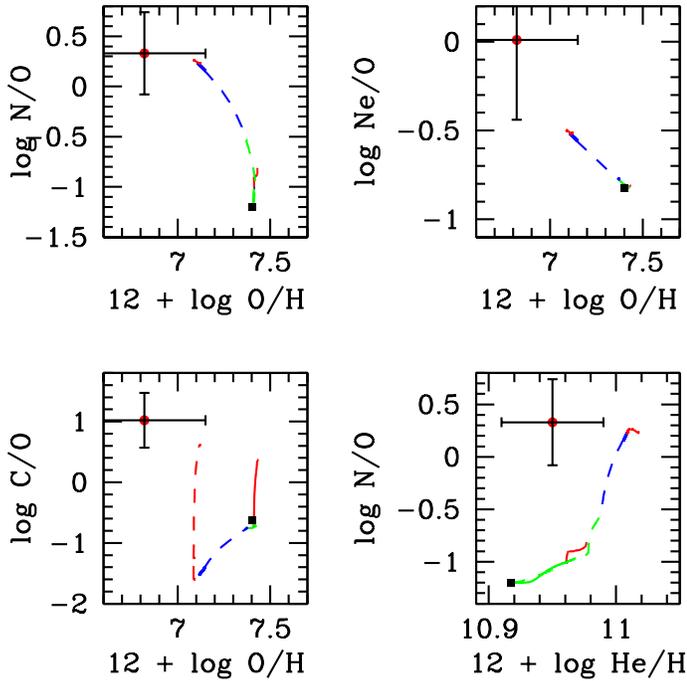} 
  \caption{Comparison of the chemical composition in \TS\ with the results
    of stellar evolution models for a star with initial mass 1\msun, and initial composition as indicated by the black dot. The model without rotation is represented in continuous line, the model with rotation is represented with dotted lines. The modifications of the chemical composition on the main sequence are indicated in green, those during 2nd dredge-up are in blue, and those corresponding to the TP-AGB phase in red. }
    \label{fig:thibaut}
\end{figure}

We note however that those models use standard assumptions. In particular
they do not account for rotation-induced mixing that is known to affect 
stellar evolution and nucleosynthesis (see \citealp{2009A&A...495..271D} and
references therein).  Additionally, the rotational transport itself may
have been strengthened by the stellar coupling in the binary system
\citep[see
  e.g.][]{2009A&A...497..243D,1997MNRAS.289..869P}. Unfortunately binary
stellar models treating the effect of mixing on the nucleosynthesis of
  low-mass stars are not yet available, but we can use
single rotating AGB models to infer the impact of mixing during the pre-AGB
phase on the surface chemical composition.

We thus compare the abundance obtained for our nebula to standard and
rotating models of a 1$M_\odot$ star computed with the code
STAREVOL (V2.92) \citep{SiessDufour2000,Siess2006} from the pre-main
sequence to the end of the TP-AGB phase (for more details, see \citealp{2009arXiv0907.5200D}). In the rotating model, an initial rotation of
100~km~s$^{-1}$ is assumed on the zero age main sequence. This is the mean value given by
  \citet{1993A&A...269..267G} for a star with the effective temperature
  of our models near the ZAMS ($\sim$7000~K). 
  This observational value is
obtained from stars in the Hyades, which are more metal-rich than our
model. If low-metallicity stars are born with the same amount of angular
momentum we could expect, due to their compactness, an even higher
initial velocity. Thus our model indicates the minimal chemical changes we
would obtained when rotation is taken into account.

The
transport of angular momentum and chemical species is driven by meridional
circulation and shear turbulence \citep{Zahn1992,MaederZahn1998}. The metallicity
is set to $Z=5 \times 10^{-4}$ (i.e., ${\rm [Fe/H]}\simeq -1.6$), which
  corresponds to the upper limit for the metallicity derived from the  Ar
  abundance in \TS{}.  The composition is scaled to the Solar one according to the
\citet{GrevesseSauval1998} mixture and enhancement in $\alpha$-elements
([$\alpha$/Fe] = +0.3 dex) is accounted for. We use the OPAL opacity
tables \citep{IglesiasRogers1996} above  a temperature of 8\,kK that account for C and O
enrichments, and the \citet{FergusonAlexander2005} data at lower
temperatures. We follow the evolution of 53 chemical species from $^1$H to
$^{37}$Cl. We use the NACRE nuclear reaction rates
\citep{AnguloArnould1999} by default and those by
\citet{CaughlanFowler1988} otherwise \citep[see][]{SiessArnould2008}. The
treatment of convection is based on the classical mixing length formalism
with $\alpha_{\rm MLT} = 1.75$, and 
diffusive
 overshoot is added below the
convective envelope according to \citet{Herwig2000}. The mass loss rate is
computed with the \citet{Reimers1975} formula (with $\eta_{\rm R} = 0.5$) for
the pre-AGB phase and then we use the \citet{VassiliadisWood1993} one during the AGB
phase. We also add diffusive overshoot following \citet{Herwig2000} to allow
third-dredge up during the AGB phase.

When comparing the results from rotating AGB models with those of non rotating ones,
we find considerable differences in the predictions of the chemical
composition as illustrated by Fig 18.  In particular rotation-induced
mixing efficiently transports chemical species from the H-burning shell to
the surface during central He-burning. This tends to decrease the surface
abundances of C and O in favour of that of N through the CNO-cycle while He
increases only mildly.  As can be seen in Fig.~\ref{fig:thibaut},
rotation-induced mixing leads to low O and high N abundances already at the
end of the second From Fig.~\ref{fig:thibaut}, we see that introducing  
  rotation (and thus strong internal mixing which can mimick binary interactions) dramatically improves  the agreement of the predicted abundances with the chemical composition of \TS. However the O depletion in the model is  consistent only with the upper limit
  allowed by the observations. The agreement would be improved if the star
  rotates faster during the phase of central H- and He-burning so that 
  more efficient mixing induces stronger O depletion and N enhancement.
  As already explained above the chosen initial velocity for our models
    could be slightly too low for its metallicity. Furthermore an even
    more pronounced rise of the rotational rate can be the result
  of the first
  mass transfer when the massive primary star expands. This high velocity can
  also lead to a larger enhancement of Ne as required by the observations.
In this way the full observational pattern of \TS\ can be explained, 
except for the
He abundance which is still higher than observed.

\section{Conclusions}
\label{sec:conclusions}
The planetary nebula \png\ (\sbs), here referred to as \TS, has been a source of many surprises.
  It is the most oxygen-poor planetary nebula
known.   Its stellar core consists of a
close binary, whose period of approximately 3.9 hours is the shortest
known of any planetary nebula nucleus.  The binary nucleus is the first
known example of a double degenerate system in a planetary nebula.
It is also the best candidate progenitor of a supernova of
type Ia.  Here, we consider the chemical composition of the nebular
shell in detail.  In a companion paper, \cite{Tovmassian09}
study the nature, history, and evolution of the binary system.

We have obtained observational data in a complete wavelength range in order to pin down the abundances of half a dozen of  elements in \TS.  The data consist of imaging and spectroscopy in the UV, optical and IR ranges. Optical data were obtained at different telescopes, resulting in high quality and reliable observational material for this object which exhibits an exceptionally line-poor optical spectrum. 

 We then defined  a strategy to determine the abundances in \TS. Since no direct electron temperature diagnostic is available, it was necessary to apply tailored photoionization modelling. We did this with much care, taking into account all the observational information at our disposal, in order to minimize the uncertainties in the derived abundances.
The photoionization modelling was performed using the pseudo-3D photoionization code Cloudy$\_$3D based on CLOUDY (version c07.02.01). Using CLOUDY, we took advantage of its detailed treatment of the hydrogen atom, which is crucial in this density-bounded nebula. Using the 3D features allowed us to take into account the apparent geometry of the nebula and the location, sizes and orientations of the observing slits, thus minimizing the errors linked to aperture effects. The spectral energy distribution of the ionizing radiation was obtained from appropriate model atmospheres.

One interesting feature of \TS\ is  that both  its stellar components  contribute to its ionization: the ``cool'' one provides the bulk of hydrogen ionization, and the ``hot'' one  is responsible for the presence of the most highly charged ions.  Thus, one can say that \TS\ has two ionizing stars. This is the first such case discovered among planetary nebulae!

The abundances of C, N, O, and Ne in the nebula are found to be respectively,  1/3.5, 1/4.2, 1/70, and   1/11 of the Solar value, with error bars of $\pm$ 0.3\,dex. For S and Ar, the abundances are less than  1/30 of  Solar. This makes of \TS\ the planetary nebula with the most extreme composition known so far. In particular, it is by far the most oxygen-deficient. The abundance of helium is  12 + log He/H = 10.95, with an accuracy that is not as high as one might have desired ($\pm$0.04 dex). The nebular abundances in \TS\ are found to be consistent with those in the atmosphere of the cool star, except for carbon which appears higher by $\sim$\,0.8\,dex than in the star.

The observed properties of the stellar core allow us to put limits on the
masses and age of the binary system, as explained in \cite{Tovmassian09} and recalled in the present paper. In particular, the initial
mass of the progenitor of the present nebula was $\sim$ 0.9\,\msun. 
We have computed state-of-the-art AGB models 
with appropriate mass and metallicity, to compare with the chemical composition of \TS. 
While a non-rotating model cannot explain the observed abundance patterns, the introduction of a 100\kms\ initial rotation in the model (which may be seen as a way to mimick the mixing due to the presence of the companion)
greatly improves the comparison. Rotation-induced mixing leads to low O and high N abundances already at the
end of the second dredge-up, whereas the C abundance increases afterwards
during each third dredge-up. 
However to fully reproduce the observed pattern requires that the
  binary interactions (and especially the first episode of mass transfer)
  spin up the star to produce a very efficient mixing. This conjecture
  still needs to be verified by proper binary models.

One feature for which we have no explanation so far is the fact that the carbon abundance in the nebula seems to be  much higher than in its progenitor. This fact would however need to be confirmed with deeper observations in the optical and the UV.

With respect to our first publication on \TS\,  by Tovmassian et al. (2001), our understanding of the status and properties of this object has thus made enormous progress. While this object is unique when considering all its observed properties together, similar objects might be plentiful in the Galactic halo --although not necessarily in a stage exhibiting a detectable planetary nebula.

\begin{acknowledgements}

This work, which extended over a period of more than seven years, could be accomplished thanks to extensive use of e-mail, short visits and informal meetings  on the occasion of conferences. We thank Gary Ferland and his associates for having offered CLOUDY to the community. We are indebted to A. Weiss for providing
unpublished evolutionary tracks.
 We are grateful to Yuri Izotov for sending the \TS\ spectrum he obtained at Kitt Peak and for drawing our attention to the existence of this object in the SDSS data base (he found it while examining by eye the hundreds of thousands of spectra from the main galaxy sample!).  Funding for the Sloan Digital Sky Survey (SDSS) has been provided by the Alfred P. Sloan Foundation, the Participating Institutions, the National Science Foundation, 
the U.S. Department of Energy, the National Aeronautics and Space Administration, the 
Japanese Monbukagakusho, the Max Planck Society, and the Higher Education Funding 
Council for England.  G. S. and S. S.-D. acknowledges the hospitality and financial support of the Instituto de Astronomia of the UNAM in Mexico City and Ensenada during many visits. C. M. acknowledges the hospitality and support of the Observatoire de Paris and the grant CONACyT-49749. G. S., C. M., R. Sz, R. N., S. S-D. acknowledge support from the European Associated Laboratory "Astrophysics Poland-France". G. T. acknowledges a grant from UC-MEXUS, which made his stay at CASS UCSD possible and 
continuous support from CONACyT. T.R. is supported by the German Federal Ministry of Education and Research (BMBF) grant 50\,OR\,0806. M. R.  acknowledges the following grants: CONACyT 43121 and 82066, DGAPA-UNAM IN116908, 112103, 108506, and 108406.
M.P. acknowledges financial support  from UNAM-DGAPA grants IN118405 and IN112708 and from a
 CONACyT-CNRS 2008 project. R.Sz. acknowledge partial support from grant N203 393334 of MNiSW/Poland. C.C. and T.D. acknowledge financial support from the french Programme National de Physique Stellaire (PNPS) of CNRS/INSU,
and from the Swiss National Science Foundation (FNS).  L. Yu. acknowledges  support
from RFBR grant 07-02-00454 and Presidium of the Russian Academy of Sciences Program ``Origin, Evolution and Structure of the Universe Objects''.  S. S.-D. acknowledges support from the Spanish Ministerio de Educaci\'on y Ciencia under the MEC/fulbright postdoctoral program.
\end{acknowledgements}

\bibliographystyle{aa}
\bibliography{stasinska}


\end{document}

%% file: SBSallfluxes-paper.txt
	 & 			 & 	 	SDSS			 & 	 	Kitt Peak			 & 	 	CFHT 2003		 	 & 		CFHT 2001	 	\\
	 & 			 & 					 & 					 & 					 & 					\\
	 & 	[Ne V]	3426	 & 					 & 					 & 		55.54	 $\pm$ 	5.19	 & 		84.05	 $\pm$ 	13.07	\\
	 & 	[O II]	3727	 & 	 				 & 	 				 & 	 $<$	0.60			 & 	 $<$ 	1.55			\\
	 & 	H I	3735	 & 					 & 					 & 		1.07	 $\pm$ 	0.31	 & 					\\
	 & 	H I	3751	 & 					 & 		4.94	 $\pm$ 	1.18	 & 		1.70	 $\pm$ 	0.32	 & 					\\
	 & 	H I	3772	 & 					 & 		4.36	 $\pm$ 	1.30	 & 		2.30	 $\pm$ 	0.35	 & 					\\
	 & 	H I	3798	 & 					 & 	 $<$	3.0		 	 & 		2.64	 $\pm$ 	0.38	 & 					\\
	 & 	H I	3836	 & 		6.73	 $\pm$ 	0.53	 & 		3.71	 $\pm$ 	0.82	 & 		4.67	 $\pm$ 	0.37	 & 					\\
	 & 	[Ne III]	3869	 & 	 $<$ 	1.2		 	 & 	 $<$	1.5			 & 		0.77	 $\pm$ 	0.31	 & 		1.67	 $\pm$ 	0.48	\\
	 & 	H I	3889	 & 		9.06	 $\pm$ 	0.43	 & 		7.81	 $\pm$ 	0.87	 & 		9.28	 $\pm$ 	0.45	 & 		2.37	 $\pm$ 	0.47	\\
	 & 	H I	3970	 & 		13.29	 $\pm$ 	0.44	 & 		13.24	 $\pm$ 	0.79	 & 		12.70	 $\pm$ 	0.47	 & 		5.85	 $\pm$ 	0.58	\\
	 & 	He II	4027	 & 	 $<$ 	1.2			 & 	 $<$	1.5			 & 		0.56	 $\pm$ 	0.21	 & 					\\
	 & 	C III	4069	 & 	 	2.1	::		 & 	 $<$	1.5			 & 	 $<$ 	0.58			 & 					\\
	 & 	H I	4102	 & 		25.11	 $\pm$ 	0.42	 & 		25.49	 $\pm$ 	0.62	 & 		25.16	 $\pm$ 	0.43	 & 		19.79	 $\pm$ 	1.06	\\
	 & 	He II	4201	 & 	 $<$ 	1.2			 & 	 $<$	1			 & 		1.04	 $\pm$ 	0.19	 & 					\\
	 & 	C II	4267	 & 	 $<$ 	1.2			 & 	 $<$	1			 & 	 $<$ 	0.32			 & 					\\
	 & 	H I	4340	 & 		46.09	 $\pm$ 	0.38	 & 		48.06	 $\pm$ 	0.44	 & 		47.19	 $\pm$ 	0.50	 & 		42.49	 $\pm$ 	0.80	\\
	 & 	[O III]	4363	 & 	 $<$ 	1.2			 & 	 $<$	1			 & 	 $<$ 	0.41		 	 & 					\\
	 & 	N III	4379	 & 	 $<$ 	1.2			 & 	 $<$	1			 & 	 $<$ 	0.09			 & 					\\
	 & 	He I	4471	 & 	 $<$ 	1.2			 & 	 $<$	1			 & 	 $<$ 	0.11			 & 					\\
	 & 	He II	4543	 & 		2.97	 $\pm$ 	0.33	 & 		2.91	 $\pm$ 	0.35	 & 		2.46	 $\pm$ 	0.22	 & 					\\
	 & 	[Ar V]	4626	 & 	 $<$ 	1.2			 & 	 $<$	1			 & 	 $<$ 	0.38			 & 					\\
	 & 	O IV	4632	 & 	 $<$ 	1.2			 & 	 $<$	1			 & 	 $<$ 	0.38			 & 					\\
	 & 	C III	4650	 & 	 $<$ 	1.2			 & 	 $<$	1			 & 	 $<$ 	0.38			 & 					\\
	 & 	C IV	4659	 & 	 $<$ 	1.2			 & 	 $<$	1			 & 	 $<$ 	0.38			 & 					\\
	 & 	C IV	4659	 & 	 $<$ 	1.2			 & 	 $<$	1			 & 	 $<$ 	0.24			 & 					\\
	 & 	He II	4686	 & 		75.13	 $\pm$ 	0.31	 & 		79.02	 $\pm$ 	0.28	 & 		77.50	 $\pm$ 	0.95	 & 		79.06	 $\pm$ 	1.11	\\
	 & 	[Ar IV]	4711	 & 	 $<$ 	1.2			 & 	 $<$	1			 & 	 $<$ 	0.15			 & 					\\
	 & 	[Ne IV]	4715	 & 	 $<$ 	1.2			 & 	 $<$	1			 & 		0.32	 $\pm$ 	0.08	 & 					\\
	 & 	[Ne IV]	4725	 & 	 $<$ 	1.2			 & 	 $<$	1			 & 		0.26	 $\pm$ 	0.08	 & 					\\
	 & 	[Ar IV]	4740	 & 	 $<$ 	1.2			 & 	 $<$	1			 & 	 $<$ 	0.15			 & 					\\
	 & 	H$\beta$	4861	 & 		100.00	 $\pm$ 	0.29	 & 		100.00	 $\pm$ 	0.26	 & 		100.00	 $\pm$ 	0.78	 & 		100.00	 $\pm$ 	1.59	\\
	 & 	O V	4930	 & 	 $<$ 	1.2			 & 	 $<$	0.8			 & 	 $<$ 	0.28			 & 					\\
	 & 	N V	4945	 & 	 $<$ 	1.2			 & 	 $<$	0.8			 & 	 $<$ 	0.28			 & 					\\
	 & 	[O III]	4959	 & 	 $<$ 	1.2			 & 	 $<$	0.99	 $\pm$ 	0.29	 & 		0.59	 $\pm$ 	0.13	 & 		0.69	 $\pm$ 	0.69	\\
	 & 	[O III]	5007	 & 		1.82	 $\pm$ 	0.28	 & 		2.16	 $\pm$ 	0.26	 & 		2.53	 $\pm$ 	0.16	 & 		1.81	 $\pm$ 	0.79	\\
	 & 	[Fe VI]	5146	 & 	 $<$ 	10.0			 & 	 $<$	0.8			 & 	 $<$ 	0.28			 & 					\\
	 & 	He II	5411	 & 		5.8	 $\pm$ 	0.22	 & 		6.49	 $\pm$ 	0.28	 & 					 & 		5.12	 $\pm$ 	0.31	\\
	 & 	He I	5876	 & 	 $<$ 	0.8			 & 	 $<$	0.8			 & 					 & 	 $<$	0.19			\\
	 & 	[Fe VII]	6087	 & 	 $<$ 	0.8			 & 	 $<$	0.8			 & 	 				 & 					\\
	 & 	H$\alpha$	6563	 & 	 	322.03	 $\pm$ 	0.15	 & 	 	306.38	 $\pm$ 	0.22	 & 	 				 & 		248.54	 $\pm$ 	2.63	\\
	 & 	[N II]	6584	 & 	 $<$ 	1.6			 & 	 				 & 	 				 & 					\\
	 & 	[S II]	6716	 & 	 $<$ 	1.6			 & 	 				 & 	 				 & 					\\
	 & 	[S II]	6731	 & 	 $<$ 	1.6			 & 	 				 & 	 				 & 					\\
	 & 	[Ar V]	7006	 & 	 $<$ 	1.6			 & 	 				 & 	 				 & 					\\
	 & 	He I	7065	 & 	 $<$ 	1.6			 & 	 				 & 	 				 & 					\\
	 & 	[Ar III]	7136	 & 	 $<$ 	1.6			 & 	 				 & 	 				 & 					\\
	 & 	[O II]	7320	 & 	 $<$ 	1.6			 & 	 				 & 					 & 					\\
	 & 	[O II]	7330	 & 	 $<$ 	1.6			 & 	 				 & 					 & 					\\
	 & 	[O II]	7333	 & 	 $<$ 	1.6			 & 	 				 & 					 & 					\\
	 & 	??	7408	 & 		0.82	 $\pm$ 	0.24	 & 	 				 & 					 & 					\\
	 & 	[Cl IV]	7530	 & 	 $<$ 	1.6			 & 	 				 & 					 & 					\\
	 & 	He II	7600	 & 	 $<$ 	1.6			 & 	 				 & 	 				 & 					\\
	 & 	O V	7611	 & 	 $<$ 	1.6			 & 	 				 & 	 				 & 					\\
	 & 	O IV	7713	 & 	 $<$ 	1.6			 & 	 				 & 	 				 & 					\\
	 & 	C IV	7726	 & 	 $<$ 	1.6			 & 	 				 & 	 				 & 					\\
	 & 	[Ar III]	7751	 & 	 $<$ 	1.6			 & 	 				 & 					 & 					\\
	 & 	[Cl IV]	8046	 & 	 $<$ 	1.6			 & 	 				 & 					 & 					\\
	 & 	CIII	8196	 & 	 $<$ 	1.6			 & 	 				 & 					 & 					\\
	 & 	He II	8237	 & 		1.76	 $\pm$ 	0.16	 & 	 				 & 					 & 					\\
	 & 	H P16	8502	 & 	 $<$ 	1.6			 & 	 				 & 					 & 					\\
	 & 	H P15	8545	 & 	 $<$ 	1.6			 & 	 				 & 					 & 					\\
	 & 	H P14	8598	 & 	 $<$ 	1.6			 & 	 				 & 					 & 					\\
	 & 	H P13	8665	 & 	 $<$ 	1.6			 & 	 				 & 					 & 					\\
	 & 	H P12	8750	 & 		1.29	 $\pm$ 	0.26	 & 	 				 & 					 & 					\\
	 & 	H P10	9014	 & 		1.7	:		 & 	 				 & 					 & 					\\
	 & 	[S III]	9069	 & 	 $<$ 	1.6			 & 	 				 & 					 & 					\\

%% file: table5-11sept-final.txt
	{}	(1)	&	(2)$^a$	& 	& (3)	&	(4)	&	(5)	&	(6)	&	(7)	&	(8)	&	(9)	\\
	{}		&	 	&		&		&		&	\textbf{R}	&	\textbf{Mi}	&	\textbf{Ma}	&	\textbf{HeMi}	&	\textbf{HeMa} 	\\
	{}		&	 	&		&		&		&		&		&		&		&	 	\\
	{}	 	 & 	 	 & 	 	 &	 {$T_h$}      	 & 	 10$^3$K 	 &	170	 &	170	 &	170	 &	170	 &	170	 \\
	{}	 	 & 	 	 & 	 	 &	 {$L_h$}      	 & 	{[\lsun]}	 &	2564	 &	1618	 &	4064	 &	1618	 &	4064	 \\
	{}	 	 & 	 	 & 	 	 &	 {log $g_h$}   	 & 	         	 &	6.7	 &	6.7	 &	6.7	 &	6.7	 &	6.7	 \\
	{}	 	 & 	 	 & 	 	 &	 {$n_0$}  	 & 	 cm$^{-3}$	 &	181	 &	181	 &	181	 &	181	 &	181	 \\
		Sun$^{b}$	&		&		&		&		&		&		&		&		&		\\
	{}	10.93	 & 		 & 	 	 &	 {He$^{c}$ }       	 & 	         	 &	10.95	 &	10.95	 &	10.95	 &	10.98	 &	10.91	 \\
	{}	8.39	 & 		 & 		 & 	 {C$^{c}$ }        	 & 		&	7.84	&	7.64	&	8.05	&	7.64	&	8.05	 \\
	{}	7.78	 & 		 & 		 & 	 {N$^{c}$ }        	 & 		&	7.15	&	7	&	7.32	&	7	&	7.32	 \\
	{}	8.66	 & 		 & 		 & 	 {O$^{c}$ }        	 & 		&	6.82	&	6.63	&	7.13	&	6.63	&	7.13	 \\
	{}	7.85	 & 		 & 	 	 &	 {Ne$^{c}$ }       	 & 	         	 &	6.83	 &	6.76	 &	6.9	 &	6.76	 &	6.9	 \\
	{}	7.14	 & 		 & 	 	 &	 {S$^{c}$ }        	 & 	         	 &	5.65	 &	5.5	 &	5.83	 &	5.5	 &	5.83	 \\
	{}	6.18	 & 		 & 	 	 &	 {Ar$^{c}$ }       	 & 	         	 &	4.7	 &	4.5	 &	4.92	 &	4.5	 &	4.92	 \\
		 	&		&		&		&		&		&		&		&		&		\\
	{}	 	 & 	 	 & 	 	 &	 {$d$}   	 & 	 {[kpc]}	 &	22.5	 &	22.3 	 &	22.9 	 &	22.3 	 &	22.9 	 \\
	{}	 	 & 	 	 & 	 	 &	 {$F_\beta^{d}$}  	 & 		 &	2.55	 &	2.52	 &	2.56	 &	2.54	 &	2.54	 \\
	{}	 	 & 	 	 & 	 	 &	 {$M$(H)} 	 & 	{[\msun]}	 &	0	 &	0.14	 &	0.15	 &	0.14	 &	0.15	 \\
	{}	 	 & 	 	 & 	 	 &	 {$f$(STIS) }   	 & 	         	 &	0.91	 &	0.91	 &	0.91	 &	0.92	 &	0.91	 \\
	{}	 	 & 	 	 & 	 	 &	 {$f$(IRS) }	 & 	         	 &	0.91 	 &	0.88	 &	0.94	 &	0.87 	 &	0.95	 \\
		 	&		&		&		&	$\Delta$$O/O$	&		&		&		&		&		\\
	{}	         H8 	 & 	0	 & 	   	 &	9.51	 & 	0.15	 &	9.01	 &	9.02	 &	9.02	 &	8.96	 &	9.09	 \\
	{}	         H8 	 & 	4	 & 	   	 &	9.29	 & 	0.15	 &	9.15	 &	9.19	 &	9.14	 &	9.12	 &	9.21	 \\
	{}	         H7 	 & 	0	 & 	   	 &	12.99	 & 	0.1	 &	14.18	 &	14.22	 &	14.18	 &	14.12	 &	14.28	 \\
	{}	         H7 	 & 	4	 & 	   	 &	13.6	 & 	0.1	 &	14.39	 &	14.46	 &	14.36	 &	14.36	 &	14.47	 \\
	{}	         H6 	 & 	0	 & 	   	 &	25.66	 & 	0.07	 &	23.64	 &	23.70	 &	23.64	 &	23.54	 &	23.81	 \\
	{}	         H6 	 & 	4	 & 	   	 &	25.71	 & 	0.07	 &	23.87	 &	23.96	 &	23.85	 &	23.80	 &	24.03	 \\
	{}	H$\gamma$	 & 	0	 & 	   	 &	47.84	 & 	0.05	 &	47.40	 &	47.47	 &	47.34	 &	47.41	 &	47.41	 \\
	{}	H$\gamma$	 & 	4	 & 	   	 &	46.73	 & 	0.05	 &	47.64	 &	47.71	 &	47.59	 &	47.65	 &	47.66	 \\
	{}	H$\alpha$	 & 	5	 & 	   	 &	251.41	 & 	0.04	 &	282.06	 &	282.48	 &	281.30	 &	283.21	 &	280.50	 \\
	{}	H$\alpha$	 & 	4	 & 	   	 &	311.03	 & 	0.04	 &	279.89	 &	280.14	 &	279.23	 &	280.85	 &	278.43	 \\
	{}	    HeI5876 	 & 	5	 & 	$<$	 &	0.19	 & 	         	 &	0.16	 &	0.25	 &	0.11	 &	0.27	 &	0.10	 \\
 	{}	    HeI7065 	 & 	4	 & 	$<$	 &	1.44	 & 	         	 &	0.04	 &	0.06	 &	0.03	 &	0.06	 &	0.02	 \\
	{}	   HeII1640 	 & 	1	 & 	   	 &	609.18	 & 	0.1	 &	594.53	 &	572.27	 &	623.3	 &	607.55	 &	575.72	 \\
	{}	   HeII4686 	 & 	0	 & 	   	 &	77.88	 & 	0.04	 &	76.53	 &	73.85	 &	79.67	 &	78.35	 &	73.60	 \\
	{}	   HeII4686 	 & 	4	 & 	   	 &	74.93	 & 	0.04	 &	76.16	 &	74.00	 &	78.91	 &	78.52	 &	72.93	 \\
	{}	   HeII5412 	 & 	4	 & 	   	 &	5.43	 & 	0.1	 &	6.19	 &	6.02	 &	6.38	 &	6.40	 &	5.88	 \\
		 	&		&		&		&		&		&		&		&		&		\\
	{}	  CIII]1909 	 & 	1	 & 	$<$	 &	61.44	 & 	         	 &	73.41	 &	89.01	 &	59.96	 &	93.05	 &	56.73	 \\
	{}	    CIII977 	 & 	2	 & 	$<$	 &	430.3	 & 	         	 &	104.58	 &	112.54	 &	96.78	 &	117.74	 &	92.22	 \\
	{}	    CIV1549 	 & 	1	 & 	   	 &	1054.84	 & 	0.2	 &	1050.89	 &	890.87	 &	1246.07	 &	916.75	 &	1200.76	 \\
	{}	    CIV4659 	 & 	0	 & 	$<$	 &	0.38	 & 	         	 &	0.14	 &	0.08	 &	0.26	 &	0.08	 &	0.26	 \\
		 	&		&		&		&		&		&		&		&		&		\\
 	{}	  [NIII]57.2 	 & 	6	 & 	 	 &	 	     & 	 	         &	1.23 &	1.54 &	0.96 &	1.56 &	0.93	 \\				
	{}	  NIII]1750 	 & 	1	 & 	$<$	 &	60.84	 & 	         	 &	3.00	 &	4.29	 &	1.99	 &	4.49	 &	1.87	 \\
	{}	    NIII991 	 & 	2	 & 	$<$	 &	423.77	 & 	         	 &	7.54	 &	9.52	 &	5.91	 &	9.81	 &	5.69	 \\
	{}	   NIV]1486 	 & 	1	 & 	   	 &	73.05	 & 	0.5	 &	65.89	 &	67.40	 &	61.95	 &	69.93	 &	58.82	 \\
	{}	     NV1240 	 & 	1	 & 	   	 &	339.3	 & 	0.3	 &	330.38	 &	253.7	 &	425.38	 &	259.25	 &	410.86	 \\
		 	&		&		&		&		&		&		&		&		&		\\
	{}	 [OIII]4363 	 & 	0	 & 	$<$	 &	0.42	 & 	         	 &	0.11	 &	0.14	 &	0.12	 &	0.15	 &	0.11	 \\
	{}	 [OIII]5007 	 & 	0	 & 	   	 &	2.52	 & 	0.3	 &	2.57	 &	3.24	 &	2.58	 &	3.35	 &	2.49	 \\
	{}	   OIV]1402 	 & 	1	 & 	$<$	 &	36.97	 & 	         	 &	7.68	 &	6.88	 &	10.60	 &	7.15	 &	10.05	 \\
	{}	 [OIV]25.9 	 & 	3	 & 	   	 &	13.72	 & 	0.3	 &	13.57	 &	11.20	 &	20.35	 &	11.24	 &	20.13	 \\
	{}	    OVI1032 	 & 	2	 & 	$<$	 &	407.42	 & 	         	 &	50.48	 &	30.45	 &	105.06	 &	30.65	 &	102.45	 \\
		 	&		&		&		&		&		&		&		&		&		\\
	{}	[NeIII]3869 	 & 	0	 & 	   	 &	0.79	 & 	0.5	 &	0.82	 &	1.45	 &	0.44	 &	1.50	 &	0.43	 \\
	{}	[NeIII]15.5 	 & 	3	 & 	$<$	 &	7.22	 & 	         	 &	0.21	 &	0.37	 &	0.11	 &	0.38	 &	0.11	 \\
	{}	[NeIII]36.0 	 & 	3	 & 	$<$	 &	28.88	 & 	         	 &	0.02	 &	0.03	 &	0.01	 &	0.03	 &	0.01	 \\
	{}	 [NeIV]2424 	 & 	6	 & 	   	 &	           	 & 	         	 &	25.20	 &	29.46	 &	20.51	 &	30.10	 &	19.93	 \\
	{}	  [NeV]3426 	 & 	0	 & 	   	 &	57.67	 & 	0.2	 &	58.32	 &	48.63	 &	66.92	 &	49.19	 &	65.84	 \\
	{}	  [NeV]14.3 	 & 	3	 & 	   	 &	29.62	 & 	0.3	 &	43.25	 &	35.05	 &	50.69	 &	34.84	 &	50.91	 \\
	{}	  [NeV]24.3 	 & 	3	 & 	   	 &	54.17	 & 	0.2	 &	54.44	 &	44.16	 &	63.76	 &	43.90	 &	64.03	 \\
	{}	  [NeVI]7.6 	 & 	6	 & 	   	 &	           	 & 	         	 &	27.47	 &	15.98	 &	45.07	 &	15.84	 &	45.48	 \\
		 	&		&		&		&		&		&		&		&		&		\\
	{}	 [ArIV]4711 	 & 	0	 & 	$<$	 &	0.15	 & 	         	 &	0.15	 &	0.15	 &	0.14	 &	0.16	 &	0.13	 \\
	{}	 [ArIV]4740 	 & 	0	 & 	$<$	 &	0.15	 & 	         	 &	0.12	 &	0.12	 &	0.11	 &	0.13	 &	0.10	 \\
	{}	  [ArV]7005 	 & 	4	 & 	$<$	 &	1.44	 & 	         	 &	0.27	 &	0.23	 &	0.31	 &	0.23	 &	0.30	 \\
	{}	   [ArV]8.0 	 & 	6	 & 	   	 &	           	 & 	         	 &	0.29	 &	0.22	 &	0.37	 &	0.22	 &	0.36	 \\
	{}	  [ArV]13.1 	 & 	3	 & 	$<$	 &	7.23	 & 	         	 &	0.45	 &	0.33	 &	0.57	 &	0.33	 &	0.57	 \\
		 	&		&		&		&		&		&		&		&		&		\\
	{}	 [SIII]9069 	 & 	4	 & 	$<$	 &	1.41	 & 	         	 &	0.24	 &	0.31	 &	0.17	 &	0.32	 &	0.16	 \\
	{}	 [SIII]18.7 	 & 	3	 & 	$<$	 &	7.22	 & 	         	 &	0.21	 &	0.25	 &	0.17	 &	0.25	 &	0.16	 \\
	{}	 [SIII]33.5 	 & 	3	 & 	$<$	 &	36.1	 & 	         	 &	0.34	 &	0.41	 &	0.28	 &	0.42	 &	0.27	 \\
	{}	  [SIV]10.5 	 & 	3	 & 	$<$	 &	7.23	 & 	         	 &	6.65	 &	6.29	 &	6.88	 &	6.35	 &	6.74	 \\
		 	&		&		&		&		&		&		&		&		&		\\
	{}	    NV1240/  NIV]1486 	 & 	 1/1 	 & 	 	 &	4.645	 & 	0.7	 &	5.01 	 &	3.76 	 &	6.86 	 &	3.71	 &	6.98 	 \\
	{}	 [OIV]25.9/[OIII]5007 	 & 	 3/0 	 & 	 	 &	0.139	 & 	0.4	 &	0.13 	 &	0.087	 &	0.20 	 &	0.085	 &	0.21	 \\
	{}	 [OIV]25.9/[OIII]5007 	 & 	 3/4 	 & 	 	 &	0.194	 & 	0.4	 &	0.22 	 &	0.136	 &	0.35 	 &	0.134	 &	0.36	 \\
	{}	 [NeV]3426/[NeIII]3869 	 & 	 0/0 	 & 	 	 &	73.021	 & 	0.7	 &	71.33	 &	33.47 	 &	151.80 	 &	32.84	 &	154.26	 \\
	{}	 [NeV]3426/[NeIV]4720 	 & 	 0/0 	 & 	 	 &	99.039	 & 	0.3	 &	127.42	 &	89.66 	 &	181.34 	 &	87.90	 &	185.72	 \\
	{}	 [NeV]24.3/ [NeV]3426 	 & 	 3/0 	 & 	 	 &	0.024	 & 	0.3	 &	0.024	 &	0.023	 &	0.024	 &	0.020	 &	0.025	 \\
	{}	 [NeV]24.3/ [NeV]14.3 	 & 	 3/3 	 & 	 	 &	1.828	 & 	0.3	 &	1.26	 &	1.26	 &	1.26	 &	1.26	 &	1.26	 \\